\documentclass[11pt]{article}
\pdfoutput=1
\usepackage[linktocpage=true]{hyperref}
\usepackage[all,knot,matrix,arrow]{xy}
\usepackage{color}
\usepackage[textwidth = 430 pt, textheight = 630 pt]{geometry}
\definecolor{MyDarkBlue}{rgb}{0.15,0.25,0.45}
\usepackage{epsfig,rotating}
\usepackage{amsmath,amssymb}
\usepackage{amsfonts}
\usepackage{mathrsfs}
\usepackage{bbm}
\usepackage{hyperref}
\hypersetup{
colorlinks=true,
citecolor=MyDarkBlue,
linkcolor=MyDarkBlue,
urlcolor=MyDarkBlue,
 pdfauthor={Patricia Ritter and Christian S\"amann},
 pdftitle={Lie 2-algebra models},
 pdfsubject={hep-th}
 breaklinks=true
 }

\let\fn\footnote
\renewcommand{\footnote}[1]{\linespread{1.1}\fn{#1}\linespread{1.29}}

\makeatletter\renewcommand{\section}{\@startsection
{section}{1}{\z@}{-3.5ex plus -1ex minus
    -.2ex}{2.3ex plus .2ex}{\bf }}
\makeatletter\renewcommand{\subsection}{\@startsection{subsection}{2}{\z@}{-3.25ex
plus -1ex minus
   -.2ex}{1.5ex plus .2ex}{\it }}
\makeatletter\renewcommand{\subsubsection}{\@startsection{subsubsection}{3}{-2.45ex}{-3.25ex
plus -1ex minus -.2ex}{1.5ex plus .2ex}{\it }}
\renewcommand{\thesection}{\arabic{section}}
\renewcommand{\thesubsection}{\arabic{section}.\arabic{subsection}}
\renewcommand{\@seccntformat}[1]{\@nameuse{the#1}.~~}

\renewcommand{\theequation}{\thesection.\arabic{equation}}
\makeatletter \@addtoreset{equation}{section}

\newcommand{\appendices}{
\section*{Appendix}\label{appendices}\setcounter{subsection}{0}
\addcontentsline{toc}{section}{Appendix}
\setcounter{equation}{0}
\makeatletter
\renewcommand{\theequation}{\Alph{subsection}.\arabic{equation}}
\renewcommand{\thesubsection}{\Alph{subsection}}
\@addtoreset{equation}{subsection}
\makeatother
}

\usepackage{epsfig,rotating}
\usepackage{amsmath,amssymb}
\usepackage{amsfonts}
\usepackage{mathrsfs}

\usepackage{graphicx}
\usepackage{xypic}

\def\slasha#1{\setbox0=\hbox{$#1$}#1\hskip-\wd0\hbox to\wd0{\hss\sl/\/\hss}}

\def\periodb#1{\setbox0=\hbox{$#1$}#1\hskip-\wd0\hbox to\wd0{-}}

\newcommand{\unit}{\mathbbm{1}}   			% identity map/matrix
\newcommand{\id}{\mathrm{id}}   			% identity map/matrix
\newcommand{\xt}{\tilde{x}}

\newcommand{\CA}{\mathcal{A}}    			% cal-letters

\newcommand{\CC}{\mathcal{C}}

\newcommand{\CF}{\mathcal{F}}

\newcommand{\CH}{\mathcal{H}}

\newcommand{\CL}{\mathcal{L}}
\newcommand{\CM}{\mathcal{M}}

\newcommand{\CN}{\mathcal{N}}
\newcommand{\CO}{\mathcal{O}}

\newcommand{\frg}{\mathfrak{g}}				% frak-letters
\newcommand{\frH}{\mathfrak{H}}

\newcommand{\FR}{\mathbbm{R}}     			% field of real numbers
\newcommand{\FC}{\mathbbm{C}}     			% field of complex numbers
\newcommand{\NN}{\mathbbm{N}}     			% set of natural numbers
\newcommand{\RZ}{\mathbbm{Z}}     			% ring of integers
\newcommand{\dd}{\mathrm{d}}     			% total differential

\newcommand{\dpar}{\partial}     			% partial differential
\newcommand{\embd}{{\hookrightarrow}}     		% embedded
\newcommand{\diag}{{\mathrm{diag}}}     		% diagonal matrix
\newcommand{\de}{\mathrm{e}}     			% Euler's number
\newcommand{\di}{\mathrm{i}}     			% imaginary unit
\newcommand{\eps}{{\varepsilon}}			% antisymmetric tensors
\newcommand{\psib}{{\bar{\psi}}}

\newcommand{\eand}{{\qquad\mbox{and}\qquad}}     		% and etc. in equations

\newcommand{\efor}{{\qquad\mbox{for}\qquad}}

\newcommand{\der}[1]{\frac{\dpar}{\dpar #1}}   		% partielle ableitung, 1 argument
\newcommand{\derr}[2]{\frac{\dpar #1}{\dpar #2}}   	% partielle ableitung, 2 argumente
\newcommand{\dderr}[2]{\frac{\dd #1}{\dd #2}}   	% totale ableitung
\newcommand{\tr}{\,\mathrm{tr}\,}     			% trace
\newcommand{\ad}{\mathrm{ad}}     			% adjoint action
\newcommand{\au}{\mathfrak{u}}
\newcommand{\asu}{\mathfrak{su}}
\newcommand{\aso}{\mathfrak{so}}

\renewcommand{\sb}{\mathsf{b}}

\newcommand{\sMat}{\mathsf{Mat}}

\newcommand{\sSO}{\mathsf{SO}}

\newcommand{\sAut}{\mathsf{Aut}\,}
\newcommand{\sEnd}{\mathsf{End}\,}
\newcommand{\sHom}{\mathsf{Hom}\,}

\newcommand{\acton}{\vartriangleright}     			% span
\newcommand{\remark}[1]{}     				% remark
\def\tyng(#1){\hbox{\tiny$\yng(#1)$}}			% small Young diagram
\def\tyoung(#1){\hbox{\tiny$\young(#1)$}}			% small Young diagram
\newcommand{\beq}{\begin{eqnarray}}
\newcommand{\eeq}{\end{eqnarray}}

\newcommand{\dotsp}{-}

\newenvironment{conditions}{
\vspace{-2mm}\begin{itemize}
  \setlength{\itemsep}{1pt}
  \setlength{\parskip}{0pt}
  \setlength{\parsep}{0pt}
}{\vspace{-2mm}\end{itemize}}

\begin{document}

\begin{titlepage}
\begin{flushright}
 EMPG--13--14
\end{flushright}
\vskip 2.0cm
\begin{center}
{\LARGE \bf Lie 2-algebra models}
\vskip 1.5cm
{\Large Patricia Ritter$^a$ and Christian S\"amann$^b$}
\setcounter{footnote}{0}
\renewcommand{\thefootnote}{\arabic{thefootnote}}
\vskip 1cm
{\em${}^a$ Centro de Estudios Cient\'ificos (CECs)\\
Avenida Arturo Prat 514, Valdivia, Chile
}\\[0.5cm]
{\em ${}^b$ Maxwell Institute for Mathematical Sciences\\
Department of Mathematics, Heriot-Watt University\\
Colin Maclaurin Building, Riccarton, Edinburgh EH14 4AS,
U.K.}\\[0.5cm]
{Email: {\ttfamily PRitter@cecs.cl , C.Saemann@hw.ac.uk}}
\end{center}
\vskip 1.0cm
\begin{center}
{\bf Abstract}
\end{center}
\begin{quote}
In this paper, we begin the study of zero-dimensional field theories with fields taking values in a semistrict Lie 2-algebra. These theories contain the IKKT matrix model and various M-brane related models as special cases. They feature solutions that can be interpreted as quantized 2-plectic manifolds. In particular, we find solutions corresponding to quantizations of $\FR^3$, $S^3$ and a five-dimensional Hpp-wave. Moreover, by expanding a certain class of Lie 2-algebra models around the solution corresponding to quantized $\FR^3$, we obtain higher BF-theory on this quantized space.
\end{quote}
\end{titlepage}

\tableofcontents

\section{Introduction and motivation}

One of the fundamental problems in theoretical physics today is the construction of theories that are formulated without reference to any specific space-time geometry. In such background independent models, space-time is expected to emerge from the dynamics of the theory, for example as vacuum configurations. A good example of such a theory is the IKKT matrix model \cite{Ishibashi:1996xs}, which was conjectured to provide a non-perturbative and background independent formulation of superstring theory. This model arises as a finite regularization of the type IIB superstring in Schild gauge. It is a zero-dimensional theory, in which fields take values in a (matrix) Lie algebra.

It has become more and more evident that many of the algebraic structures underlying string and M-theory are not Lie algebras but rather extensions of Lie algebras which are known as strong homotopy Lie algebras or $L_\infty$-algebras. In particular, regularizations of the membrane action yield models with fields taking values in truncated, 2-term $L_\infty$-algebras. It is therefore desirable to study generalized IKKT-like models, in which fields can take values in strong homotopy Lie algebras. The purpose of this paper is to initiate such a study. 

To keep our models manageable, we will restrict ourselves to the 2-term strong homotopy Lie algebras, which are categorically equivalent to semistrict Lie 2-algebras\footnote{In this paper, we will use the terms ``Lie 2-algebra'' and ``2-vector spaces'' rather freely. Unless stated otherwise, we will use them to refer to 2-term strong homotopy Lie algebras and 2-term chain complexes of vector spaces, respectively.}. These algebras feature prominently in higher gauge theories which seem to underlie M-brane models, and a subclass of these form the gauge structure of the recently popular M2-brane models \cite{Bagger:2007jr,Gustavsson:2007vu,Aharony:2008ug}. This is to be seen in analogy to the conventional Lie algebras of the IKKT model underlying the gauge theories arising in the effective description of D-brane configurations in string theory.

This paper is structured as follows. In the remainder of this section,
we will give a more detailed motivation for studying Lie 2-algebra
models. We then review relevant definitions of Lie 2-algebras and
discuss various notions of inner products on them in section
2. Section 3 makes contact with the quantization of 2-plectic
manifolds. Homogeneous and inhomogeneous Lie 2-algebra models are then discussed in section 4 and section 5, respectively. We present our conclusions in section 6. Two appendices summarize useful definitions and review the gauge symmetry of semistrict higher gauge theory for the reader's convenience.

\subsection{Background independence and the IKKT model}\label{ssec:background_IKKT}

As stated above, it is an important goal to construct and study background independent theories to replace our mostly background dependent formulations of string theory. A straightforward method for eliminating the background geometry from any field theory is to dimensionally reduce it to a point. If the fields in the original theory took values in a Lie algebra and its adjoint representation, one is left with a matrix model.

Matrix models have indeed contributed greatly to the understanding of non-perturba\-tive phenomena in string theory. This started with the Hermitian matrix models describing $c<1$ string theory \cite{DiFrancesco:1993nw} and continued with the success of the IKKT matrix model \cite{Ishibashi:1996xs}, see also \cite{Aoki:1998bq}.

The IKKT model is obtained by regularizing the Green-Schwarz action of the type IIB superstring in Schild gauge,
\begin{equation}
 S=\int \dd^2\sigma \sqrt{g} \alpha \left(\tfrac{1}{4}\{X_\mu,X_\nu\}^2-\tfrac{\di}{2}\psib \Gamma^\mu\{X_\mu,\psi\}\right)+\beta\sqrt{g}~.
\end{equation}
In this regularization, the worldsheet fields $X_\mu$ and $\psi$ are replaced by hermitian matrices $A_\mu$ and $\psi$, while the integral becomes a trace and the Poisson bracket $\{\dotsp,\dotsp\}$ is turned into the commutator $-\di [\dotsp,\dotsp]$. Note that this process is standard in noncommutative field theory and the result is the following:
\begin{equation}\label{eq:action_IKKT}
 S_{\rm IKKT}=\alpha\tr\left(-\tfrac{1}{4}\,[A_\mu,A_\nu]^2-\tfrac{1}{2}\,\psib\Gamma^\mu[A_\mu,\psi]+\beta \unit\right)~.
\end{equation}
Alternatively, one can obtain the IKKT model by dimensionally reducing maximally supersymmetric Yang-Mills theory in ten dimensions to a point. The fields $A_\mu$ and $\psi$ here take values in the gauge algebra of the ten-dimensional theory.

As equations of motion of the action \eqref{eq:action_IKKT}, we have
\begin{equation}\label{eq:eom_IKKT}
\begin{aligned}
 {}[A_\mu,[A^\mu,A^\nu]]-\tfrac{\di}{2}\Gamma^\nu_{\alpha\beta}\{\psi^\beta,\psib^\alpha\}&=0~,\\
 \Gamma^\mu_{\alpha\beta}[A_\mu,\psi^\beta]&=0~.
\end{aligned}
\end{equation}
Amongst the solutions to these equations are matrices $A_m$,
$m=1,\ldots,2d$, that we can identify with the generators $\hat{x}^m$ of the Heisenberg algebra $[\hat{x}^m,\hat{x}^n]=\di \theta^{mn}\unit$. The generators $\hat{x}^m$ are the coordinate functions on the Moyal space $\FR^{2d}_\theta$, and this is the most prominent example of a geometry emerging as the vacuum configuration of the IKKT model. Expanding the action \eqref{eq:action_IKKT} around this background solution as $A_m=\hat{x}_m+\hat{A}_m$, we obtain Yang-Mills theory on noncommutative $\FR^{2d}_\theta$ \cite{Aoki:1999vr}. The action \eqref{eq:action_IKKT} therefore simultaneously provides the background and the dynamics of the theory.

More general noncommutative geometries are obtained as vacuum solutions of deformations of the IKKT model. A particularly interesting class of deformations comprise mass-terms as well as a cubic potential term, 
\begin{equation}\label{eq:action_IKKT_def}
 S_{\rm def}=S_{\rm IKKT}+\tr\left( -\tfrac{1}{2}\sum_\mu m^2_{1,\mu}A_\mu A_\mu+\tfrac{\di}{2}m_2 \psib \psi+c_{\mu\nu\kappa}A^\mu A^\nu A^\kappa\right)~,
\end{equation}
where $c_{\mu\nu\kappa}$ is some background tensor field, cf.\ \cite{Berenstein:2002jq}. This action has classical configurations corresponding to fuzzy spheres and the space $\FR^3_\lambda$, which is a discrete foliation of $\FR^3$ by fuzzy spheres, as well as noncommutative Hpp waves, see \cite{DeBellis:2010sy} and references therein. 

Finally, note that in a very similar manner in which a background expansion of the IKKT model yields Yang-Mills theories on noncommutative spaces, one can also obtain models of gravity, see e.g.\ \cite{Steinacker:1003.4134}.

\subsection{Lie $n$-algebras in string theory}

Lie $2$-algebras arise in the categorification of the notion of a Lie
algebra. In this process, the vector space underlying the Lie algebra is replaced by a category. Furthermore, the standard structural equations of a Lie algebra, which state that the Lie bracket is antisymmetric and
satisfies a Jacobi identity, are lifted in a controlled way and hold
only up to an isomorphism in this category. Lie $n$-algebras arise analogously by $n$-fold, iterative categorification of Lie algebras. In the semistrict case, which is the one we will consider exclusively in this paper, Lie $n$-algebras are equivalent to truncated $n$-term strong homotopy Lie or $L_\infty$-algebras, which are also known as $L_n$-algebras.

Strong homotopy Lie algebras and in particular their truncated versions appear in a variety of contexts related to string theory, for example:
\begin{itemize}
\setlength{\itemsep}{-1mm}
\item[$\triangleright$] Strong homotopy Lie algebras arise in string field theory, cf.\ \cite{Zwiebach:1992ie,Kajiura:0410291}, as well as in Kontsevich's deformation quantization. 
\item[$\triangleright$] Lie 2-algebras appear in topological open M2-brane actions in the form of Courant Lie 2-algebroids \cite{Hofman:2002rv}.
\item[$\triangleright$] Special Lie 2-algebras, which are known as differential crossed modules, form the gauge structure of the recently popular M2-brane models \cite{Bagger:2007jr,Gustavsson:2007vu,Aharony:2008ug} as shown in \cite{Palmer:2012ya}.
\item[$\triangleright$] The full M2-brane action is coupled to the $C$-field of supergravity and is thus expected to be related to parallel transport of two-dimensional objects, which has an underlying Lie 3-algebra \cite{Sati:0801.3480}.
\item[$\triangleright$] Interactions of M5-branes are mediated by M2-branes ending on them and their boundaries are one-dimensional objects known as self-dual strings. It is natural to assume that an effective description of M5-branes yields a higher gauge theory describing the parallel transport of these self-dual strings. The gauge structure of such a higher gauge theory is described by a Lie 2-algebra, cf.\ \cite{Baez:2010ya}.
\item[$\triangleright$] Equations of motion of interacting non-abelian superconformal field theories in six dimensions have been derived using twistor spaces in \cite{Saemann:2012uq,Saemann:2013pca}. These constructions again make use of the framework of higher gauge theory, employing Lie 2- and 3-algebras.
\end{itemize}

\subsection{Our goals in this paper}

We saw above that the Lie algebras describing gauge symmetries in effective descriptions of D-branes within string theory are replaced by Lie 2-algebras in M-theory. It is therefore natural to suspect that a potential non-perturbative description of M-theory along the lines of the IKKT model may be based on Lie 2-algebras.

In this paper, we perform an initial study of zero-dimensional field theories in which the fields take values in a Lie 2-algebra. We discuss the mathematical notions required in the description of Lie 2-algebra models, put them into context and test how far the analogies with the IKKT model reach. 

Throughout this paper, we will distinguish two types of models. First, there are {\em homogeneous Lie 2-algebra models}, in which the fields $\{X^a\}$ take values in the direct sum of the two vector spaces $V$ and $W$ that underlie a Lie 2-algebra. In the {\em inhomogeneous models}, we have two types of fields $\{X^a\}$ and $\{Y^i\}$, where the $X^a$ take values in $V$ while the $Y^i$ take values in $W$. Note that homogeneous models form a subset of the inhomogeneous models.

Since semistrict Lie 2-algebras contain ordinary Lie algebras, homogeneous Lie 2-algebra models will trivially contain the IKKT model as a special case. Moreover, Lie 2-algebras contain the 3-algebras appearing in M2-brane models, and we therefore also expect our Lie 2-algebra models to contain the 3-algebra models discussed previously in \cite{Sato:2009mf,Lee:2009ue,Sato:2009tr,Sato:2010ca}, \cite{Furuuchi:2009ax} and \cite{DeBellis:2010sy}. 

In \cite{Sato:2009mf}, the author followed the logic of the IKKT model, starting from a Schild-type action of the M2-brane \cite{Park:2008qe},
\begin{equation}
 S=T_{\rm M2} \int \dd^3\sigma \{X^M,X^N,X^K\}^2~,~~~M,N,K=0,\ldots,10~.
\end{equation}
He then suggested to regularize this action by replacing the Nambu-bracket by that of a 3-Lie algebra. Note that it has often been suggested that, at quantum level, Nambu-Poisson structures should turn into 3-Lie algebras, see \cite{DeBellis:2010pf} and references therein. To a certain extent, one can even make the resulting action supersymmetric\footnote{Full supersymmetry, however, seems to be possible only for four scalar fields with a metric of split signature \cite{Furuuchi:2009ax}.}, and the result is \cite{Furuuchi:2009ax,Sato:2009tr}
\begin{equation}\label{eq:3LA_action}
 S_{\rm 3LA}=\langle [X^M,X^N,X^K],[X^M,X^N,X^K]\rangle+\langle \bar{\Psi},\Gamma_{MN} [X^M,X^N,\Psi]\rangle~,
\end{equation}
where $\Psi$ is a Majorana spinor of $\sSO(1,10)$. A very similar model has been studied in \cite{Chu:2011yd} as a matrix model for the description of multiple M5-branes.

Alternatively, one can obtain a zero-dimensional action with fields living in a 3-Lie algebra by dimensionally reducing the M2-brane models to a point. The case of the BLG-model was discussed in \cite{DeBellis:2010sy}, where various solutions have been interpreted as quantized Nambu-Poisson manifolds. Compared to \eqref{eq:3LA_action}, there are additional scalar fields present, living in the inner derivations of the underlying 3-Lie algebra that arise from the dimensional reduction of the Chern-Simons part and the covariant couplings to the matter fields. While there is now a dichotomy of fields compared to \eqref{eq:3LA_action}, the resulting action is invariant under 16 supercharges. Moreover, applying a dimensionally reduced form of the Higgs mechanism proposed in \cite{Mukhi:2008ux}, this action reduces to \eqref{eq:action_IKKT_def} in the strong coupling limit as shown in \cite{DeBellis:2010sy}.

An important feature of the IKKT model is that familiar examples of quantized symplectic manifolds arise as solutions of the classical equations of motion. Correspondingly, we expect that ``higher quantized'' manifolds arise as solutions of our Lie 2-algebra models. There are essentially two approaches in the literature of how to extend geometric quantization to a higher setting. First, we can focus on the Poisson structure and generalize this structure to a Nambu-Poisson structure. The geometric quantization of Nambu-Poisson manifolds, however, is problematic and the answers obtained in this context are not very satisfying, see \cite{DeBellis:2010pf} and references therein. The second approach focuses on extending the symplectic structure to a 2-plectic one, which yields a Lie 2-algebra of Hamiltonian 1-forms on the manifold. This is by now a fairly standard construction in multisymplectic geometry \cite{Cantrijn:1999aa,Baez:2008bu}. Quantizing the 2-plectic manifold amounts here to quantizing the Lie 2-algebra of Hamiltonian 1-forms. While more appealing than the first one, this approach has its own shortcomings, and a more detailed discussion is found in section \ref{ssec:2-plectic_quantization}. Here it is important to note that this point of view is clearly very suitable for our purposes, and we expect that quantized versions of Lie 2-algebras of Hamiltonian 1-forms yield solutions to the classical equations of motion of our Lie 2-algebra models.

From this perspective, our Lie 2-algebra models are a good testing ground for the extension of the notion of a space. In noncommutative geometry, the first step in such an extension is made by replacing the commutative product in the algebra of functions by a noncommutative one. The next step is to generalize this to a nonassociative product, which requires the use of 2-term $L_\infty$- and $A_\infty$-algebras. Ultimately, the notion of a commutative algebra of functions on a manifold should be generalized to that of a certain type of operad or an even more general mathematical structure.

\section{Lie 2-algebras}

Lie 2-algebras are categorified versions of Lie algebras. While categorification is not a unique or straightforward recipe, the procedure is roughly the following: most mathematical notions are based on spaces endowed with extra structure satisfying certain basic equations. To categorify such a notion, replace the spaces with categories and endow them with extra structure given by functors that satisfy the basic equations up to an isomorphism. The isomorphisms, in turn, have to satisfy reasonable coherence equations. In the case of Lie algebras, one thus obtains the {\em weak Lie 2-algebras} \cite{Roytenberg:0712.3461}: the linear space underlying the Lie algebra gets replaced by a linear category. We demand that we have a Lie bracket functor on this category, but it is antisymmetric and satisfies the Jacobi identity only up to isomorphisms. These isomorphisms are called the alternator and the Jacobiator, respectively.

Demanding that the alternator is trivial, which implies that the categorified Lie bracket is antisymmetric, one obtains the so-called {\em semistrict Lie 2-algebras}. It is these that we will be considering in this paper. They are particularly nice to work with, as they are categorically equivalent to 2-term $L_\infty$-algebras, cf.\ \cite{Baez:2003aa}. 

One can go one step further and demand that the Jacobi identity is satisfied, too. In this case, one ends up with {\em strict Lie 2-algebras}, which can be identified with differential crossed modules \cite{Baez:2002jn}. Although most of the structural generalizations of categories have been lost at this point, strict Lie 2-algebras are still interesting. For example, they underlie the definition of non-abelian gerbes, see e.g.\ \cite{Baez:2010ya}. Moreover, when endowed with a metric, they contain all the 3-algebras that have appeared recently in M2-brane models \cite{Palmer:2012ya}.

\subsection{Semistrict Lie 2-algebras}

As stated above, semistrict Lie 2-algebras are categorically equivalent to 2-term $L_\infty$-algebras, and we can relatively easily specify their structure in terms of vector spaces. The general definition of an $L_\infty$-algebra is recalled for the reader's convenience in appendix \ref{app:A}.

A {\em 2-term $L_\infty$-algebra} is given by a two-term complex of real\footnote{To simplify the notation for inner products later on, we restrict ourselves to real vector spaces.} vector spaces,
\begin{equation}
  V~\xrightarrow{~\mu_1~}~W~\xrightarrow{~\mu_1~}~0~,
\end{equation}
where gradings $-1$ and $0$ are assigned to elements of $V$ and $W$, respectively. This complex is equipped with unary, binary and ternary totally graded antisymmetric and multilinear ``products'' $\mu_1$, $\mu_2$ and $\mu_3$ satisfying the following higher homotopy relations:
\begin{subequations}\label{eq:homotopy_relations}
\begin{equation}
\begin{aligned}
 \mu_1(w)&=0~,~~~\mu_2(v_1,v_2)=0~,\\
 \mu_1(\mu_2(w,v))&=\mu_2(w,\mu_1(v))~,~~~\mu_2(\mu_1(v_1),v_2)=\mu_2(v_1,\mu_1(v_2))~,\\
 \mu_3(v_1,v_2,v_3)&=\mu_3(v_1,v_2,w)=\mu_3(v_1,w_1,w_2)=0~,\\
 \mu_1(\mu_3(w_1,w_2,w_3))&=-\mu_2(\mu_2(w_1,w_2),w_3)-\mu_2(\mu_2(w_3,w_1),w_2)-\mu_2(\mu_2(w_2,w_3),w_1)~,\\
 \mu_3(\mu_1(v),w_1,w_2)&=-\mu_2(\mu_2(w_1,w_2),v)-\mu_2(\mu_2(v,w_1),w_2)-\mu_2(\mu_2(w_2,v),w_1) 
\end{aligned}
\end{equation}
and
\begin{equation}
\begin{aligned}
 \mu_2(\mu_3(w_1,&w_2,w_3),w_4)-\mu_2(\mu_3(w_4,w_1,w_2),w_3)+\mu_2(\mu_3(w_3,w_4,w_1),w_2)\\
 & -\mu_2(\mu_3(w_2,w_3,w_4),w_1)=\\
 &\mu_3(\mu_2(w_1,w_2),w_3,w_4)-\mu_3(\mu_2(w_2,w_3),w_4,w_1)+\mu_3(\mu_2(w_3,w_4),w_1,w_2)\\
 &-\mu_3(\mu_2(w_4,w_1),w_2,w_3)
 -\mu_3(\mu_2(w_1,w_3),w_2,w_4)-\mu_3(\mu_2(w_2,w_4),w_1,w_3)~,
\end{aligned}
\end{equation}
\end{subequations}
where $v,v_i\in V$ and $w,w_i\in W$.

Besides the above product, we also introduce the product $\kappa_2:V\times V\rightarrow V$ with
\begin{equation}
 \kappa_2(v_1,v_2):=\mu_2(\mu_1(v_1),v_2)=-\mu_2(\mu_1(v_2),v_1)=-\kappa_2(v_2,v_1)~.
\end{equation}

A simple example of a semistrict Lie 2-algebra is the following one \cite{Baez:2003aa}, which we will denote by $(\frg,V,\rho,c)$: as two-term complex, we take $V\rightarrow \frg$, where $\frg$ is a finite-dimensional real Lie algebra and $V$ is a vector space carrying a representation $\rho$ of $\frg$. The non-vanishing products are given by
\begin{equation}
 \mu_2(g_1,g_2):=[g_1,g_2]~,~~~\mu_2(g,v)=-\mu_2(v,g):=\rho(g)v~,~~~\mu_3(g_1,g_2,g_3)=c(g_1,g_2,g_3)~,
\end{equation}
where $g\in \frg$, $v\in V$ and $c\in H^3(\frg,V)$. Since $\mu_1$ is trivial, isomorphic objects in the category corresponding to this Lie 2-algebra are identical. Such Lie 2-algebras are called {\em skeletal}.

Any semistrict Lie 2-algebra is in fact categorically equivalent to a skeletal one, and all skeletal semistrict Lie 2-algebras are equivalent to one of the form $(\frg,V,\rho,c)$ \cite[Thm. 55]{Baez:2003aa}. This fact can be used to classify Lie 2-algebras.

If $V=\FR$ then an interesting example of a Lie-algebra cocycle is given by $c(g_1,g_2,g_3)=k\langle g_1,[g_2,g_3]\rangle$, where $\langle\dotsp,\dotsp\rangle$ is the Killing form and $k\in \FR$. The resulting semistrict Lie 2-algebra is also called the string Lie 2-algebra of $\frg$.

Other examples are given by the Lie 2-algebra of Hamiltonian 1-forms on 2-plectic manifolds, and we describe these in detail in section \ref{ssec:2Plectic_Manifolds}.

If the Jacobiator $\mu_3$ in a semistrict Lie 2-algebra vanishes, we arrive at a {\em strict Lie 2-algebra} or, equivalently, a {\em differential crossed module}: both $\mu_2$ and $\kappa_2$ now satisfy the Jacobi identity, and we have a two-term complex of Lie algebras $V\xrightarrow{\mu_1}W$ with an action $\acton W\times V\rightarrow V: w\acton v:=\mu_2(w,v)$ satisfying
\begin{equation}
 \mu_1(w\acton v)=[w,\mu_1(v)]\eand \mu_1(v_1)\acton v_2=[v_1,v_2]~,
\end{equation}
for all $v\in V$ and $w\in W$, where the commutators are identified with $\mu_2$ on $W$ and $\kappa_2$ on $V$.

The simplest examples of strict Lie 2-algebras are the gauge algebras of $\au(1)$-bundles and $\au(1)$-gerbes: the Lie algebra $\au(1)$ can be regarded as a Lie 2-algebra\footnote{Here and in the following, $*$ denotes the trivial Lie algebra $\{0\}$.} $(*\xrightarrow{\mu_1} \au(1),\acton)$, where $\mu_1(*)=0\in \au(1)$ and $\acton$ is trivial. The gauge algebra of a $\au(1)$-gerbe is the Lie 2-algebra $\sb\au(1)=(\au(1)\xrightarrow{\mu_1} *,\acton)$, where $\mu_1$ and $\acton$ are trivial. A non-abelian example is the derivation Lie 2-algebra $\mathsf{Der}(\frg)$ of a Lie algebra $\frg$, $(\frg\xrightarrow{\rm ad}\mathsf{der}(\frg),\acton)$, where $\mathsf{der}(\frg)$ are the derivations of the Lie algebra $\frg$, $\ad$ is the embedding of $\frg$ as inner derivations via the adjoint map, and $\acton$ is the natural action of derivations of $\frg$ onto $\frg$.

\subsection{Lie 2-algebra homomorphisms}\label{ssec:Lie_2_algebra_homomorphism}

To analyze symmetries in our models, we will require the notion of a homomorphism between Lie 2-algebras. Such a homomorphism should preserve both the vector space structure as well as the higher products. However, as we are working in a categorified setting, we will require the higher products to be preserved only up to an isomorphism. The appropriate definition for Lie 2-algebras has been developed in \cite[Def.\ 23]{Baez:2003aa}. Translated to the equivalent 2-term $L_\infty$-algebras, we have the following definition \cite[Def.\ 34]{Baez:2003aa}.

An $L_\infty$-homomorphisms $\Psi:\,L\rightarrow L'$ between two
2-term $L_\infty$-algebras $L=V\rightarrow W$ and $L'=V'\rightarrow W'$ is defined as a set of maps
\begin{equation}
  \label{eq:55}
  \Psi_{-1}:\, V\rightarrow V'~,~~~\Psi_0:\, W\rightarrow W'~,~~~\Psi_2:\, W\times W\rightarrow V'~,
\end{equation}
where $\Psi_{-1}$ and $\Psi_0$ form a linear chain map and $\Psi_2$ is a skew-symmetric bilinear map preserving the higher product structure. That is, for $w,\,w_i\in W$
and $v,\,v_i\in V$, the following hold:
\begin{equation}\label{eq:Lie_2_algebra_homomorphism_conditions}
\begin{array}{rcl}
 \Psi_0\left(\mu_2(w_1,w_2)\right)&=&\mu_2\left(\Psi_0(w_1),\Psi_0(w_2)\right)+\mu_{1}(\Psi_2(w_1,w_2))~,\\ 
\Psi_{-1}\left(\mu_2(w,v)\right)&=&\mu_2(\Psi_0(w),\Psi_{-1}(v))+\Psi_2(w,\mu_{1}(v))~,\\
\mu_3(\Psi_0(w_1),\Psi_0(w_2),\Psi_0(w_3))&=&\Psi_{-1}\left(\mu_3(w_1,w_2,w_3)\right)-\left[
  \Psi_2(w_1,\mu_2(w_2,w_3))\right. \\
\ & \ &\left.+\mu_2\left(\Psi_0(w_1),\Psi_2(w_2,w_3)\right)+\text{cyclic}\,(w_1,w_2,w_3)\right].
\end{array}
\end{equation}
Two homomorphisms $\Psi:\,L\rightarrow L'$ and
$\Phi:\,L'\rightarrow L''$ can be combined via the composition rules
\begin{subequations}\label{composition}
\begin{align}
  \label{eq:56}
  (\Psi\circ\Phi)_0(w)&=\Psi_0\Phi_0(w)~,\\
 (\Psi\circ\Phi)_{-1}(v)&=\Psi_{-1}\Phi_{-1}(v)~,\\
 (\Psi\circ\Phi)_2(w_1,w_2)&=\Psi_{-1}\Phi_2(w_1,w_2)+\Psi_2\left(\Phi_0(w_1),\Phi_0(w_2)\right)~.
\end{align}
\end{subequations}
The identity automorphism $\text{Id}_L:\,L\rightarrow L$ is given by the maps
\begin{equation}
  \label{eq:58}
  (\text{id}_L)_0(w)=w~,~~~(\text{id}_L)_{-1}=v~,~~~(\text{id}_L)_2(w_1,w_2)=0~.
\end{equation}
The inverse to an automorphism $\Phi:L\rightarrow L$ under the composition $\circ$ given in \eqref{composition} is
indicated by $\Phi^{-1\circ}: L\rightarrow L$. It satisfies
$\Phi^{-1\circ}\circ\Phi=\Phi\circ\Phi^{-1\circ}=\text{id}_L$, and is
made of three maps given by
\begin{subequations}
\begin{align}
  \label{eq:59}
  (\Phi^{-1\circ})_0(w)&=(\Phi_0)^{-1}(w)~,\\
(\Phi^{-1\circ})_{-1}(v)&=(\Phi_{-1})^{-1}(v)~,\\
(\Phi^{-1\circ})_2(w_1,w_2)&=-(\Phi_{-1})^{-1}\left(\Phi_2\left((\Phi_0)^{-1}(w_1),(\Phi_0)^{-1}(w_2)\right)\right)~.
\end{align}
\end{subequations}
For more details on morphisms between semistrict Lie 2-algebras see for instance \cite{Baez:2003aa,Zucchini:2011aa}.

\subsection{Inner products on semistrict Lie 2-algebras}\label{ssec:inner_products}

Let us now discuss the notion of an inner product on semistrict Lie 2-algebras, which we will need to write down action functionals. Naturally, an inner product on a semistrict Lie 2-algebra should originate from an inner product on its underlying Baez-Crans 2-vector space. Moreover, it should be compatible with certain actions of Lie 2-algebra homomorphisms. And finally, as we want to be able to reproduce dimensionally reduced M2-brane models, we allow for indefinite scalar products, cf.\ appendix \ref{app:A}. 

Unfortunately, there are at least three different notions of inner product that satisfy these properties. First, there is a scalar product on $L_\infty$-algebras\footnote{This definition of a scalar product extends to other $\infty$-algebras. Moreover, it corresponds to the notion of a binary invariant polynomial of the $L_\infty$-algebra.} that was used in \cite{Zwiebach:1992ie} and \cite{Kontsevich:1992aa}, see also \cite{Igusa:2003yg} and \cite{0821843621}. Given an $L_\infty$-algebra $L=\oplus_i L_i$, a {\em (cyclic) scalar product} $\langle \dotsp,\dotsp\rangle_\infty$ on $L$ is a
non-degenerate, even, bilinear form that is compatible with all the homotopy products $\mu_n$, $n\in\NN^*$. Explicitly, we have
\begin{equation}
 \begin{aligned}
  \langle x_1,x_2\rangle_\infty&=(-1)^{\xt_1+\xt_2}\langle x_2,x_1\rangle_\infty~.\\
  \langle \mu_n(x_1,\ldots,x_n),x_0\rangle_\infty&=(-1)^{n+\xt_0(\xt_1+\cdots+\xt_n)}\langle \mu_n(x_0,\ldots,x_{n-1}),x_n\rangle_\infty~,
 \end{aligned}
\end{equation}
$x_i\in L$. Adapted to 2-term $L_\infty$-algebras, it follows that a {\em cyclic scalar product} on a semistrict Lie 2-algebra
$V\longrightarrow W$ is a scalar product $\langle \dotsp,\dotsp \rangle_\infty$
on $V\oplus W$, which satisfies the following conditions.
\begin{conditions}
\item[(i)] It is even symmetric, that is:
\begin{equation}
\langle v_1,v_2\rangle_\infty=\langle v_2,v_1\rangle_\infty~,~~~\langle w_1,w_2\rangle_\infty=\langle w_2,w_1\rangle_\infty~,~~~\langle v,w\rangle_\infty =\langle w,v\rangle_\infty=0~.
\end{equation}
\item[(ii)] It is cyclically graded symmetric with respect to $\mu_2$ and cyclically graded antisymmetric with respect to $\mu_1$ and $\mu_3$, which implies
\begin{equation}\label{eq:too_strong}
 \langle \mu_1(v),w\rangle_\infty=\langle \mu_2(v_1,w),v_2\rangle_\infty=\langle \mu_3(w_1,w_2,w_3),v\rangle_\infty=0~.
\end{equation}
\end{conditions}
We thus see that this kind of inner product is very restrictive.

Another metric $\langle \dotsp,\dotsp\rangle_{\rm red}$ was introduced on {\em reduced} semistrict Lie 2-algebras, where $\mu_1$ is injective \cite{Zucchini:2011aa}. In this case, $V$ can be regarded as a subspace of $W$ and the domain and range of all products collapse to $W$. One can then impose the following invariance conditions
\begin{equation}
 \begin{aligned}
  \langle\mu_2(w_1,w_2),w_3\rangle_{\rm red}+\langle w_2,\mu_2(w_1,w_3)\rangle_{\rm red}=0~,\\
  \langle\mu_1(\mu_3(w_1,w_2,w_3)),w_4\rangle_{\rm red}+\langle w_3,\mu_1(\mu_3(w_1,w_2,w_4))\rangle_{\rm red}&=0~.
 \end{aligned}
\end{equation}
While the latter equation is very reminiscent of the fundamental identity for 3-Lie algebras, cf.\ appendix \ref{app:A}, focusing on reduced semistrict Lie 2-algebras is a severe restriction. In particular, it excludes the semistrict (and strict) Lie 2-algebra $\sb\au(1)=\au(1)\rightarrow *$, which is the gauge 2-algebra of an abelian gerbe. Moreover, it will collide with the semistrict Lie 2-algebra structures obtained on 2-plectic manifolds in section \ref{ssec:2Plectic_Manifolds}. 

The final metric we want to consider arises from extending the definition on strict Lie 2-algebras to the semistrict case, cf.\ e.g.\ \cite{Baez:2002jn,Martins:2010ry,Palmer:2012ya}. On a semistrict Lie 2-algebra $V\oplus W$, an inner product is an even and graded symmetric bilinear map $\langle\dotsp,\dotsp \rangle_0$ such that
\begin{equation}\label{eq:15}
\begin{aligned}
  \langle v_1,v_2\rangle_0=\langle v_2,v_1\rangle_0~,~~~\langle w_1,w_2\rangle_0=\langle w_2,w_1\rangle_0~,~~~&\langle v,w\rangle_0=\langle w,v\rangle_0=0~,\\
 \langle \mu_2(w_1,x_1),x_2\rangle_0+\langle x_1,\mu_2(w_1,x_2)\rangle_0&=0
\end{aligned}
\end{equation}
for all $v_i\in V$, $w_i\in W$ and $x_i\in V\oplus W$. We will call
this inner product the {\em minimally invariant} inner product. Note
that demanding $\langle \mu_2(x_3,x_1),x_2\rangle_0+\langle
x_1,\mu_2(x_3,x_2)\rangle_0=0$ in general is too restrictive, as this
would imply that $\mu_2(v_1,w_1)=0$ due to $\langle
\mu_2(v_1,w_1),v_2\rangle_0+\langle w_1,\mu_2(v_1,v_2)\rangle_0=0$. Note furthermore that the above relations automatically imply that
\begin{equation}
 \langle \kappa_2(x_1,x_2),x_3\rangle_0+\langle x_2,\kappa_2(x_1,x_3)\rangle_0=0~.
\end{equation}

Besides matching the natural definition of an inner product on differential crossed modules, this definition includes also natural inner products on the semistrict Lie 2-algebras $(\frg,V,\rho,c)$ if $\frg$ is the Lie algebra of metric preserving transformations on $V$. And finally, it will turn out to match the natural metrics on semistrict Lie 2-algebras arising from 2-plectic manifolds.

\subsection{Transposed products}

To facilitate computations with metric semistrict Lie 2-algebras, it is useful to introduce ``transposed products'' $\mu_n^*$ for each $\mu_n$. These products are defined by regarding the products as operators acting on the element in the last slot and taking the dual:
\begin{equation}
  \begin{aligned}
    \label{eq:1}
    \langle \mu_1(y_1),y_2\rangle&=:\langle y_1,\mu_1^*(y_2)\rangle~,\\
    \langle \mu_2(x_1,y_1),y_2\rangle&=:\langle y_1,\mu_2^*(x_1,y_2)\rangle~,\\
    \langle \kappa_2(x_1,y_1),y_2\rangle&=:\langle y_1,\kappa_2^*(x_1,y_2)\rangle~,\\
    \langle \mu_3(x_1,x_2,y_1),y_2\rangle&=:\langle y_1,\mu_3^*(x_1,x_2,y_2)\rangle
  \end{aligned}
\end{equation}
for all $x_i,y_i\in V\oplus W$. The product $\mu_1^*$ had already been introduced in \cite{Baez:2002jn} and used extensively in \cite{Palmer:2012ya}. Note that $\mu_3^*$ is not antisymmetric. 

Let us examine the transposed products for each of the inner products in more detail. First, in the case of $\infty$-metrics $\langle \dotsp,\dotsp\rangle_\infty$, the only non-vanishing transposed product is
\begin{equation}
 \mu_2^*(w_1,w_2)=-\mu_2(w_1,w_2)~.
\end{equation}

In the case of metric $\langle \dotsp,\dotsp\rangle_{\rm red}$, we have 
\begin{equation}
 \mu_2(w_1,w_2)=-\mu_2^*(w_1,w_2)\eand \kappa^*_2(w_1,w_2)=-\kappa_2(w_1,w_2)
\end{equation}
for all $w_1,w_2\in W$. These are the only two transposed products that
are needed, since here one really only has to deal with the metric on $W$. 

For the metric $\langle \dotsp,\dotsp\rangle_0$, we have more generally
\begin{equation}
 \mu_2(w,x)=-\mu_2^*(w,x)
\end{equation}
for all $w\in W$ and $x\in V\oplus W$, which, together with $\kappa^*_2(x,y)=\mu_2^*(\mu_1(x),y)$, implies that
\begin{equation}
\kappa^*_2(v_1,v_2)=-\kappa_2(v_1,v_2)~. 
\end{equation}

The transposed products that cannot be reduced to the products in the Lie 2-algebra are
\begin{equation}
 \mu_1^*:W\rightarrow V~,~~~\mu_2^*:V\times V\rightarrow W\eand\mu_3^*:W\times W\times V\rightarrow W~,
\end{equation}
which have degrees $-1$, $2$ and $1$, respectively. They are defined implicitly via 
\begin{equation}
\begin{aligned}
\langle \mu_1(v_1),w_1 \rangle_0 =: \langle v_1,\mu_2^*(w_1)\rangle_0~,&~~~
   \langle \mu_2(v_1,w),v_2 \rangle_0 =: \langle w,\mu_2^*(v_1,v_2)\rangle_0~,\\
  \langle \mu_3(w_1,w_2,w_3),v \rangle_0 &=: \langle w_3,\mu_3^*(w_1,w_2,v)\rangle_0~.
\end{aligned}
\end{equation}
To simplify notation, we will only denote these three with a star from here on.

Combining our definitions with the homotopy algebra relations, we obtain the following set of equalities:
\begin{equation}
  \label{eq:starred relations 1}
  \begin{aligned}
\mu_2(\mu_1(v_1),v_2)&=\mu_2(v_1,\mu_1(v_2))=\mu_1^*(\mu_2^*(v_2,v_1))~,\\
\mu_2(\mu_1(v),w)&=\mu_1(\mu_2(v,w))=\mu_2^*(\mu_1^*(w),v)~,\\ 
\mu_1^*(\mu_2(w_1,w_2))&=\mu_2(\mu_1^*(w_1),w_2)=\mu_2(\mu_1^*(w_2),w_1)~,\\
\mu_1^*(\mu_3^*(w_1,w_2,v))&=-\mu_3(\mu_1(v),w_1,w_2)~,\\
\mu_1(\mu_3(w_1,w_2,w_3))&=-\mu_3^*(w_1,w_2,\mu_1^*(w_3))~,\\
\mu_3^*(\mu_1(v_1),w,v_2)&=-\mu_3^*(\mu_1(v_2),w,v_1)~, \\ 
\mu_3^*(\mu_1(v_1),w,v_2)&=\mu_2^*(v_1,\mu_2(w,v_2))-\mu_2^*(v_2,\mu_2(w,v_1))-\mu_2(w,\mu_2^*(v_1,v_2)~,\\
\end{aligned}
\end{equation}
as well as 
\begin{equation}
  \label{eq:19}\begin{array}{rl}
  \mu_2(w_1,\mu_3^*(w_2,w_3,v))&\!\!\!+\mu_2(w_3,\mu_3^*(w_1,w_2,v))+\mu_2(w_2,\mu_3^*(w_3,w_1,v))=\\
&\hspace{-15ex}\mu_3^*(\mu_2(w_1,w_2),w_3,v)+\mu_3^*(\mu_2(w_3,w_1),w_2,v)+\mu_3^*(\mu_2(w_2,w_3),w_1,v)\\
&\hspace{-17ex}+\mu_3^*(w_1,w_2,\mu_2(w_3,v))+\mu_3^*(w_3,w_1,\mu_2(w_2,v))+\mu_3^*(w_2,w_3,\mu_2(w_1,v))\\
&\hspace{-17ex}-\mu_2^*(\mu_3(w_1,w_2,w_3),v)~.
\end{array}
\end{equation}

\subsection{M2-brane model 3-algebras}\label{ssec:M2-brane_3_algebras}

The currently most successful M2-brane models \cite{Bagger:2007jr,Gustavsson:2007vu,Aharony:2008ug} are given by Chern-Simons matter theories, in which the gauge structure is described by a 3-algebra\footnote{See appendix \ref{app:A} for the relevant definitions.}. Note that we will use the term {\em 3-algebra} to collectively describe both the real 3-algebras of \cite{Cherkis:2008qr} and the hermitian 3-algebras of \cite{Bagger:2008se} in this paper. These 3-algebras have nothing to do with Lie 3-algebras or other categorifications of the notion of a Lie algebra. Instead, these 3-algebras are readily shown to be equivalent to certain classes of metric differential crossed modules \cite{Palmer:2012ya}. As we want to identify 3-algebra models in our Lie 2-algebra models later, let us briefly recall this construction.

We start from a strict Lie 2-algebra $L$ endowed with an inner product $\langle\dotsp,\dotsp\rangle_0$ for which $W=\frg$ is a real Lie algebra and $V$ is a vector space carrying a faithful orthogonal representation of $\frg$. The only non-trivial products are $\mu_2:W\times W\rightarrow W$ and $\mu_2:W\times V\rightarrow W$, which are given by the Lie bracket and the representation of $W$ as endomorphism on $V$, respectively.

As shown in \cite{deMedeiros:2008zh}, isomorphism classes of such data are in one-to-one correspondence to isomorphism classes of real 3-algebras. In particular, we can define implicitly an operator $D:V\times V\rightarrow W$ via
\begin{equation}
 \langle w, D(v_1,v_2)\rangle_0:=\langle \mu_2(w,v_1),v_2 \rangle_0~.
\end{equation}
With our above definitions, it follows that $D(v_1,v_2)=-\mu_2^*(v_1,v_2)$. Note that $\mu_2^*(v_1,v_2)$ is antisymmetric. We can then introduce a triple bracket $[\dotsp,\dotsp,\dotsp]:V^{\wedge 2}\times V\rightarrow V$ by
\begin{equation}\label{eq:triple_bracket}
 [v_1,v_2,v_3]:=D(v_1,v_2)\acton v_3=-\mu_2(\mu_2^*(v_1,v_2),v_3)~.
\end{equation}
This bracket satisfies by definition the fundamental identity, cf.\ \eqref{eq:FI_algebra}, and we therefore arrive at a real 3-algebra. Note that a similar construction exists for hermitian 3-algebras.
 
As the triple bracket \eqref{eq:triple_bracket} can be defined for any
Lie 2-algebra with inner product $\langle\dotsp,\dotsp\rangle_0$, one
can now ask under which condition the fundamental identity is
satisfied and the triple bracket yields a real 3-algebra. A short
computation reveals that this is only the case for arbitrary strict or
skeletal metric Lie 2-algebras. 

While there is no connection between the ternary bracket of a 3-Lie algebra and the Jacobiator of a Lie 2-algebra in general, we can construct (at least) one example where they can be essentially identified. Consider the vector space of $n\times n$ matrices $\sMat(n)$. Together with the 3-bracket
\begin{equation}
 [a,b,c]=\tr(a)[b,c]+\tr(b)[c,a]+\tr(c)[a,b]~,
\end{equation}
$\sMat(n)$ forms a 3-Lie algebra as shown in \cite{Awata:1999dz}. There, this 3-Lie algebra was suggested to appear in the quantization of Nambu-Poisson brackets. Interestingly, we can also identify this bracket with the Jacobiator of a reduced Lie 2-algebra $V\rightarrow W$, where $V=W=\sMat(n)$ and the following higher products are non-vanishing:
\begin{equation}
 \begin{aligned}
  \mu_1(v)&=v~,\\
  \mu_2(w_1,w_2)&=\tr(w_1)w_2-\tr(w_2)w_1+[w_1,w_2]~,\\
  \mu_2(v,w)&=-(\tr(v)w-\tr(w)v+[v,w])~,\\
  \mu_3(w_1,w_2,w_3)&=\tr(w_1)[w_2,w_3]+\tr(w_2)[w_3,w_1]+\tr(w_3)[w_1,w_2]
 \end{aligned}
\end{equation}
for all $v\in V$ and $w\in W$. The higher homotopy relations \eqref{eq:homotopy_relations} are readily verified. We will denote this Lie 2-algebra by $2\sMat(n)$.

\section{Quantized symplectic and 2-plectic manifolds}

Before coming to physical models, we will briefly review the quantization\footnote{In this paper, we will use a very rough notion of quantization that is sufficient for our considerations. For a more detailed discussion, see e.g.\ \cite{DeBellis:2010pf} and references therein.} of symplectic spaces and discuss generalizations of this to 2-plectic manifolds. The quantized spaces we introduce here will arise as solutions in our Lie 2-algebra models later on.

\subsection{Quantization of symplectic manifolds}

We start from a symplectic manifold $(M,\omega)$, which is regarded as the phase space of a classical mechanical system. The observables of this system are given by the functions on $M$, which form a commutative algebra under pointwise multiplication. In addition, the symplectic form induces a Lie algebra structure on the vector space of smooth functions on $M$, which turns $M$ into a Poisson manifold. Explicitly, we have for each function $f\in \CC^\infty(M)$ a corresponding Hamiltonian vector field $X_f$ defined according to $\iota_{X_f}\omega=\dd f$. The Poisson bracket on $\CC^\infty(M)$ induced by $\omega$ is then given by 
\begin{equation}
 \{f,g\}:=\iota_{X_f}\iota_{X_g}\omega~,
\end{equation}
and we denote the resulting Poisson algebra by $\Pi_{M,\omega}$.  As examples, consider $\FR^2$ and $S^2$. On these spaces the symplectic form is the volume form ${\rm vol}$ and the induced Poisson bracket in some coordinates $x^a$, $a=1,2$, reads as
\begin{equation}
 \{f_1,f_2\}=\frac{\eps^{ab}}{|{\rm vol}|} \derr{f_1}{x^a}\derr{f_2}{x^b}~.
\end{equation}

The quantization of a symplectic manifold is given by a Hilbert space $\CH$ together with a linear map $\hat{-}:\CC^\infty(M)\rightarrow \sEnd(\CH)$ such that the Poisson algebra $\Pi_{M,\omega}$ is mapped to the Lie algebra $\sEnd(\CH)$ at least to lowest order in some deformation parameter $\hbar$:
\begin{equation}\label{eq:correspondence_principle}
 [\hat{f},\hat{g}]=\hat{f}\hat{g}-\hat{g}\hat{f}=\widehat{-\di\hbar~\{f,g\}}+\CO(\hbar^2)
\end{equation}
for all $f,g\in\CC^\infty(M)$. Equation \eqref{eq:correspondence_principle} is known as the {\em correspondence principle}.

\subsection{2-plectic manifolds}\label{ssec:2Plectic_Manifolds}

Consider a smooth manifold $M$ endowed with a 3-form $\varpi$ that is
closed and non-degenerate in the sense that $\iota_X\varpi=0$ implies
$X=0$. We call such a 3-form a {\em 2-plectic form} and say that $M$
is a {\em 2-plectic manifold}. This can be regarded as a categorification of the notion of a symplectic structure. In particular, three-dimensional manifolds with volume forms $\varpi$ are 2-plectic manifolds.

While a symplectic structure on a manifold $M$ always gives rise to a Poisson structure on $M$ by taking its inverse, a 2-plectic form $\varpi$ gives rise to a {\em Nambu-Poisson structure}\footnote{See appendix \ref{app:A} for a definition and more details.} only under certain conditions \cite{springerlink:10.1007/BF00400143}. Therefore, a different analogy should be considered here.

Having discussed categorifications of Lie algebras before, it is natural to expect that there is a categorification of the Poisson algebra in terms of a semistrict Lie 2-algebra \cite{Baez:2008bu}. Define the set of {\em Hamiltonian 1-forms} $\frH(M)$ as those forms $\alpha$ for which there is a vector field $X_\alpha$ such that $\iota_{X_\alpha}\varpi=-\dd \alpha$. Note that for a three-dimensional manifold $M$, $\frH(M)=\Omega^1(M)$. We then define the semistrict Lie 2-algebra $\Pi_{M,\varpi}$ as the vector space $V\oplus W:=\CC^\infty(M)\oplus \frH(M)$ with non-vanishing products
\begin{equation}
 \pi_1(f)=\dd f~,~~~\pi_2(\alpha,\beta)=-\iota_{X_{\alpha}}\iota_{X_{\beta}}\varpi~,~~~\pi_3(\alpha,\beta,\gamma)=-\iota_{X_{\alpha}}\iota_{X_{\beta}}\iota_{X_{\gamma}}\varpi~,
\end{equation}
where $f\in\CC^\infty(M)$ and $\alpha,\beta,\gamma\in\Omega^1(M)$. Note that the bracket $\pi_2$ is Hamiltonian. That is,
\begin{equation}
 X_{\pi_2(\alpha,\beta)}=[X_\alpha,X_\beta]~,
\end{equation}
where the bracket on the right-hand side is the commutator of vector fields. Another useful identity for computations with Hamiltonian vector fields is
\begin{equation}
\iota_{[X_\alpha,X_\beta]}=\CL_{X_\alpha}\iota_{X_\beta}-\iota_{X_\beta}\CL_{X_\alpha}~. 
\end{equation}

A long-standing open question in this context is how to define the analogue of the commutative algebra of observables that on symplectic manifolds was given by the pointwise product of functions on phase space. Ordinary Poisson algebras containing both Lie and associative structure are encoded in a Poisson Lie algebroid. The higher analogue of this structure has been shown to be a so-called Courant Lie 2-algebroid, see \cite{Rogers:2010ac,Fiorenza:1304.6292} for more details on this point. To our knowledge, however, an explicit product on $\frH(M)$ has not been constructed so far. A solution to this problem might be to switch from the semistrict Lie 2-algebra $\Pi_{M,\varpi}$ to the categorically equivalent, skeletal Lie 2-algebra. Here, the 1-forms form an ordinary Lie algebra, and, if we were able to identify this Lie algebra with a matrix algebra, we could use the ordinary matrix product as a product between observables. Another solution might originate from a comparison with the loop space quantization, cf.\ \cite{Saemann:2012ab}. For our purposes, this product is not relevant, and we merely assume that it makes sense to identify observables on 2-plectic manifolds with the vector spaces underlying $\Pi_{M,\varpi}$.

From now on, let us restrict our considerations to three-dimensional Riemannian manifolds $M$ for which $\varpi$ is the volume form. We can endow the Lie 2-algebra $\Pi_{M,\varpi}$ with a metric, following the rules and definitions used in section \ref{ssec:inner_products}. For the two vector subspaces $\CC^\infty(M)$ and $\frH(M)$, we use the usual integrals with respect to the volume form $\varpi$:
\begin{equation}
  \label{eq:8}
  \langle f, g\rangle_0 := \int_{M}\varpi~f\cdot  g \eand \langle \alpha, \beta\rangle_0 := \int_{M}\alpha\wedge \star \beta~,
\end{equation}
which can be easily checked to be invariant under the action of $\pi_2(\alpha,\dotsp)$. Note that in the non-compact case, finiteness of these integrals becomes an issue. In particular, one should either restrict to classes of functions and 1-forms with finite norm or consider closed subsets of $M$ as integration domain. If possible, one might also consider a 2-plectic form $\varpi$ with appropriate fall-off behavior towards infinity. To avoid boundary contributions, we will always imply a restriction of $\Pi_{M,\varpi}$ to elements with finite norm.

Via the metric, we can now introduce the transposed product $\pi_1^*$ and $\pi_3^*$:
\begin{equation} \label{eq:11}
 \langle\pi_1(f),\alpha\rangle_0:=\langle f,\pi_1^*(\alpha)\rangle_0\eand\langle \pi_3(\alpha, \beta, \gamma), f\rangle_0:= \langle \gamma,\pi_3^*(\alpha,\beta, f)\rangle_0~,
\end{equation}
which are therefore given by
\begin{equation}
  \label{eq:12}
  \pi_1^*(\alpha)=-\star\dd \star\alpha\eand\pi_3^*(\alpha,\beta,f)=\star \,\dd\, \iota_{X_\beta}\iota_{X_\alpha} \star f~.
\end{equation}
Note that, by the non-degeneracy of $\varpi$, all combinations of products 
\begin{equation}
  \label{eq:34}
  \pi_2(\pi_1(f),\alpha)~,~~~\pi_3(\pi_1(f),\alpha,\beta)\eand\pi_3^*(\pi_1(f),\alpha,g)
\end{equation}
are identically zero, as well as 2 and 3-products containing more than
one $\pi_1(f)$, as easily derived from \eqref{eq:starred relations 1}.

\subsection{Examples}\label{ssec:2-plectic_examples}

Let us now review the manifolds $\FR^3$ and $S^3$ and their Lie 2-algebras $\Pi_{M,\varpi}$, which will appear in the analysis of the solutions of our model later on.

\paragraph{Euclidean space $\mathbb{R}^3$.} We endow three-dimensional Euclidean space $\mathbb{R}^3$ with its canonical volume form $\varpi=\tfrac{1}{3!}\eps_{ijk}\dd x^i\wedge\dd x^j\wedge\dd x^k$ written in standard Cartesian coordinates $x^i$. All 1-forms are Hamiltonian, and we compute their Hamiltonian vector fields to be
\begin{equation}
 X_\alpha=X^i_\alpha \dpar_i=-(\eps^{ijk}\dpar_j\alpha_k)\dpar_i\efor \alpha=\alpha_i\dd x^i~,
\end{equation}
which leads to the following products:
\begin{equation}
\begin{aligned}\label{eq:R^3_products}
 &\pi_1(f):=\dd f~,~~\pi_1(\alpha)\stackrel{!}{:=} 0~,\\
 &\pi_2(\alpha,\beta):=\eps^{ijk}\dpar_i\alpha_k(\dpar_j\beta_\ell-\dpar_\ell\beta_j)\dd x^\ell~,\\
 &\pi_3(\alpha,\beta,\gamma):=\eps^{ijk}\eps^{mnp}\dpar_m\alpha_n\dpar_j\beta_k(\dpar_i\gamma_p-\dpar_p\gamma_i)~.
\end{aligned}
\end{equation}

The subset of Hamiltonian 1-forms that are constant or linear\footnote{i.e.\ linear with respect to translations on $\FR^3$} together with the set of constant and linear functions and the above defined
non-trivial products $\pi_1$ and $\pi_3$ form a {\em Heisenberg Lie
  2-algebra}, the appropriate categorification of the Heisenberg
algebra. Note that higher brackets vanish on constant and exact 1-forms. The remaining linear 1-forms are given by 
\begin{equation}
  \xi_i=\tfrac{1}{2}\eps_{ijk}x^j\dd x^k~, 
\end{equation}
whose Hamiltonian vector fields are $-\dpar_i$ and for which we have 
\begin{equation}\label{eq:R^3_special_products}
 \pi_2(\xi_i,\xi_j)= \eps_{ijk}\dd x^k\eand\pi_3(\xi_i,\xi_j,\xi_k)=-\eps_{ijk}~.
\end{equation}

Note that the 1-forms $\xi_i$ have a special meaning once they are transgressed to loop space. Here, the direction given by the $\dd x^k$ is interpreted as the tangent to the loop, and one arrives at the following functions on loop space:
\begin{equation}
 \tfrac{1}{2}\eps_{ijk}\oint \dd\tau~ x^j(\tau)~ \dderr{x^k(\tau)}{\tau}~,
\end{equation}
where $\tau\in S^1$ is the loop parameter. For more details about these functions on loop space, see \cite{Palmer:2011vx,Saemann:2012ab}.

Assuming finiteness of the norm of the involved functions and 1-forms, we have the following formulas for the transpose product $\pi_3^*$:
\begin{equation}\label{eq:R^3_transposed_product}
\begin{aligned}
\pi_3^*(\alpha,\beta,f)&=-\tfrac{1}{4}\eps^{ij\ell}\eps^{mnp}\dpar_m\alpha_n(\dpar_p\beta_\ell-\dpar_\ell\beta_p)\dpar_jf\dd x_i~,\\
\pi_3^*(\xi_i,\xi_j,f)&=-\left(\partial_i f\dd x_j- \partial_j
  f\dd x_i\right)~.
\end{aligned}
\end{equation}

\paragraph{The sphere $S^3$.} The other example we are interested in is the 3-sphere $S^3$. It will turn out convenient to work in {\em Hopf coordinates} $0\leq\eta\leq\tfrac{\pi}{2}$ and
$0\leq\theta_i\leq 2\pi$, which parametrize the embedding $S^3\embd \FC^2$ via
\begin{equation}
 z_1=\de^{\di \theta_1}\sin\eta\eand z_2=\de^{\di \theta_2}\cos\eta~.
\end{equation}
Note that instead of using the standard range given above, we can also use $0\leq\eta\leq\pi$,
$0\leq\theta_1\leq 2\pi$ and $0\leq\theta_2\leq \pi$.

For simplicity, we combine them as $(\eta_1,\eta_2,\eta_3)=(\eta,\theta_1,\theta_2)$. The volume form and the metric read as
\begin{equation}
\varpi=\sin\eta_1\cos\eta_1\dd\eta_1\wedge\dd\eta_2\wedge\dd\eta_3\eand\dd s^2=\dd\eta_1^2+\sin^2\eta_1\,\dd\eta_2^2+\cos^2\eta_1\,\dd\eta_3^2~.
\end{equation}
For 1-forms $\alpha\in\Omega^1(S^3)$, we compute the following Hamiltonian vector fields 
\begin{equation}
 X_\alpha=X^i_\alpha \dpar_i=-\frac{1}{\sin\eta_1\cos\eta_1}(\eps^{ijk}\dpar_j\alpha_k)\dpar_i\efor \alpha=\alpha_i\dd \eta_i~,
\end{equation}
where now $\dpar_i:=\der{\eta_i}$. One readily derives the products:
\begin{equation}
\begin{aligned}\label{eq:S^3_products}
 &\pi_1(f):=\dd f~,~~\pi_1(\alpha)\stackrel{!}{:=} 0~,\\
 &\pi_2(\alpha,\beta):=\frac{1}{\sin\eta_1\cos\eta_1}\eps^{ijk}\dpar_i\alpha_k(\dpar_j\beta_\ell-\dpar_\ell\beta_j)\dd \eta^\ell~,\\
 &\pi_3(\alpha,\beta,\gamma):=\frac{1}{\sin^2\eta_1\cos^2\eta_1}\eps^{ijk}\eps^{mnp}\dpar_m\alpha_n\dpar_j\beta_k(\dpar_i\gamma_p-\dpar_p\gamma_i)~.
\end{aligned}
\end{equation}

Here, it is not possible to derive 1-forms from the vector fields $X_{\dpar_i}$, as $\iota_{X_{\dpar_1}}\varpi$ is not closed, and therefore it cannot equal $\dd\xi_1$. Instead, we choose the same vector fields as for $\FR^3$, corrected by a factor of $\frac{1}{\sin \eta_1 \cos\eta_1}$. This yields the 1-forms
\begin{equation}
  \xi_i=\tfrac{1}{2}\eps_{ijk}\eta^j\dd \eta^k~, 
\end{equation}
together with the following formulas for the products:
\begin{equation}
 \pi_2(\xi_i,\xi_j)=\frac{\eps_{ijk}\dd\eta^k}{\sin \eta_1 \cos\eta_1}\eand\pi_3(\xi_i,\xi_j,\xi_k)=-\frac{\eps_{ijk}\dd\eta^k}{\sin^2 \eta_1 \cos^2\eta_1}~.
\end{equation}

The formulas for the transposed product $\pi_3^*$ read as
\begin{equation}\label{eq:S^3_transposed_products}
\begin{aligned}
\pi_3^*(\alpha,\beta,f)&=-\frac{\eps^{ij\ell}\eps^{mnp}}{4\sin \eta_1 \cos\eta_1}\dpar_m\alpha_n(\dpar_p\beta_\ell-\dpar_\ell\beta_p)\dpar_jf\dd x_i~,\\
\pi_3^*(\xi_i,\xi_j,f)&=-\frac{1}{\sin \eta_1 \cos\eta_1}\left(\partial_i f\dd \eta_j- \partial_j
  f\dd \eta_i\right)~.
\end{aligned}
\end{equation}

\subsection{Reduction of 2-plectic to symplectic manifolds}\label{ssec:reduction}

The 2-plectic manifolds we will discuss appear very naturally in the context of M-theory. Roughly speaking, the 2-plectic structure on these spaces arises here as the ``dual'' of a tri-vector field originating from a non-trivial $C$-field in M-theory, cf.\ e.g.\ \cite{Chu:2009iv}. This is the higher analogue of a symplectic structure arising as a dual to the Seiberg-Witten bivector field \cite{Seiberg:1999vs}. Our 2-plectic manifolds can be seen as M-theory lifts of symplectic manifolds appearing in string theory. In the following, we briefly comment on taking the inverse of this lift.

To reduce from M-theory to type IIA string theory, we have to identify an M-theory direction along which the 2-plectic form is invariant. Instead of restricting to the usual Kaluza-Klein procedure, we should also allow non-trivial fibrations of the 2-plectic manifold over a symplectic manifold. Since we are mostly interested in three-dimensional spaces, we can regard them as contact manifolds, and, upon reducing along the Reeb vector field corresponding to the contact form, we necessarily obtain a symplectic manifold. In this process, we contract the Hamiltonian 1-forms with the Reeb vector to obtain the Poisson algebra of functions on the underlying symplectic manifold. We will discuss this reduction explicitly for $\FR^3$ and $S^3$ in the following.

Another possibility of interpreting this reduction is a slight detour via loop spaces, see e.g.\ \cite{Saemann:2012ex,Saemann:2012ab}: while the boundary of a string on a D-brane yields a point, that of an M2-brane on an M5-brane forms a loop. It is therefore naturally to consider loop spaces of the worldvolume of the M5-brane or submanifolds thereof. Switching to loop space allows us to introduce the so-called transgression map, which reduces the form degree by one: each loop comes with a natural tangent vector, which is given by the loop of the tangent vectors to the loop. Contracting an $n$-form on a manifold with this vector yields an $n-1$-form on loop space. Since this transgression map is a chain map\footnote{i.e.\ it maps closed/exact forms to closed/exact forms}, a 2-plectic form $\varpi$ on a manifold $M$ is mapped to a symplectic form on the corresponding loop space. 

To reduce the M-theory loop space to an ordinary space of string theory amounts to restricting to loops that are parallel to the Reeb vector field. Integrating over the loop parameter reduces the dependence of functions on loop space to that of the zero mode of the loop. Therefore, functions on loop space are reduced to functions on the symplectic manifold. Further support of this point of view comes from the observation that the Lie 2-algebra $\Pi_{M,\varpi}$ transgresses to a Poisson algebra on the loop space of $M$. The quantization of $\Pi_{M,\varpi}$ should similarly correspond to a natural quantization of the Poisson algebra on loop space, cf.\ \cite{Saemann:2012ab}.

Let us now think of the above three-dimensional spaces as contact manifolds. We want to reduce them along the Reeb vectors corresponding to a chosen contact 1-form to obtain two-dimensional manifolds. These manifolds will be endowed with a natural symplectic structure, which is given by the total derivative of the contact 1-form, restricted to the kernel of the same 1-form. Explicitly, after identifying a maximally non-integrable
1-form $\gamma$, which amounts to $\gamma\wedge\dd\gamma$ being nowhere vanishing, we need
to find the corresponding Reeb vector field $X_R$ satisfying
\begin{equation}
  \label{eq:2}
  \iota_{X_R}\gamma=1\eand\iota_{X_R}\dd\gamma=0~.
\end{equation}
Since we are working with three-dimensional manifolds, we can normalize the contact form by imposing the additional condition
\begin{equation}
 \gamma\wedge \dd \gamma=\varpi~.
\end{equation}
Now, every 1-form $\gamma$ in $\Pi_{M,\varpi}$ has its
corresponding Hamiltonian vector field $X_\gamma$, and we have also
$\dd\gamma=-\iota_{X_\gamma}\varpi$, so that
$\iota_{X_\gamma}\dd\gamma=0$. That is, $X_\gamma$ satisfies the
second requirement of a Reeb vector. Moreover,
$\iota_{X_\gamma}\gamma=-1$ since
\begin{equation}
  \label{eq:4}
  0\neq \dd \gamma=-\iota_{X_\gamma}\varpi=-\iota_{X_\gamma}(\gamma\wedge \dd \gamma)=-(\iota_{X_\gamma}\gamma)\dd \gamma~.
\end{equation}
We can therefore take $X_R:=-X_\gamma$ as the Reeb vector corresponding to the contact 1-form $\gamma$. In the M-theory context, the Reeb vector field is a vector field along the `M-theory direction'.

The reduction of the 2-plectic manifold together with its induced Lie 2-algebra $\Pi_{M,\varpi}$ to a symplectic manifold with its corresponding Poisson algebra is rather straightforward: all forms are contracted by the Reeb vector field. In particular, we obtain a two-dimensional manifold\footnote{In the cases that we are interested in, the quotient space turns out to be a smooth manifold.} $M_R:=M/X_R$, where we divide $M$ by the free abelian action of the Reeb vector field. The symplectic form on $M_R$ is given by $\varpi_R:=\iota_{X_R}\varpi=\dd \gamma$. Moreover, the Hamiltonian 1-forms $\alpha$ on $M$ become functions $f_\alpha:=\iota_{X_R}\alpha$ on $M_R$ and the Lie 2-algebra $\Pi_{M,\varpi}$ reduces to a Poisson algebra $\Pi_{M_R,\varpi_R}$. Hamiltonian 1-forms along (the M-theory direction) $X_R$ are of the form $\alpha=f_\alpha \gamma$. For two such relative forms $\alpha$ and $\beta$, we have 
\begin{equation}\label{eq:3}
  \iota_{X_R}\pi_2(\alpha,\beta)=-\iota_{X_R}\iota_{X_\alpha}\iota_{X_\beta}\varpi=-\iota_{X_\alpha}\iota_{X_\beta}\varpi_R=-\iota_{X_{f_\alpha}}\iota_{X_{f_\beta}}\varpi_R=\{f_{\alpha},f_{\beta}\}~,
\end{equation}
where the Hamiltonian vector fields of the functions $f_\alpha$ are defined with respect to $\varpi_R$. Writing $\dd_R$ for the exterior derivative on $M_R$, we have
\begin{equation}
  \label{eq:5}
  \dd_R (\iota_{X_R}\alpha)=\dd_R f_{\alpha}=\iota_{X_{f_{\alpha}}}\varpi_R~.
\end{equation}
Altogether, we recover a two-dimensional symplectic manifold, with all its
structure given in terms of our initial 2-plectic one.

\paragraph{Reduction of $\mathbb{R}^3$.} To reduce the 2-plectic space $\mathbb{R}^3$ to the symplectic manifold $\FR^2$, we use the contact form $\gamma=\dd z- y\dd x$. The corresponding Reeb vector $X_R$, given by $\dd\gamma=\iota_{X_R}\varpi$, is therefore $X_R=\partial_z$. Restricting to Hamiltonian 1-forms along the M-theory direction $\gamma$, we recover the usual Poisson  algebra for $\FR^2$. Consider two such forms $\alpha=f_\alpha\gamma$ and $\beta=f_\beta\gamma$. We have
\begin{equation}\label{eq:9}
\begin{aligned}
  \iota_{\partial_z}\pi_2(\alpha,\beta)= \left\{
    \iota_{X_R}\alpha,\iota_{X_R}\beta\right\}=&\left\{f_\alpha,f_\beta\right\}=-\iota_{X_{f_\alpha}}\iota_{X_{f_\beta}}\varpi_R\\
\
=&\der{x} f_\alpha~\der{y} f_\beta-\der{y} f_\alpha~ \der{x} f_\beta~.
\end{aligned}
\end{equation}

We can also reduce the 2-plectic manifold $\FR^3\backslash\{0\}$ to $S^2$, recovering the symplectic structure there. The contact form here is given in canonical spherical coordinates by $\gamma=r^2\dd r-\cos\theta \dd \phi$. This yields the Reeb vector field $X_R=\frac{1}{r^2}\dpar_r$. The 2-plectic structure $\varpi$ reduces to the usual symplectic structure of the 2-sphere: $\varpi_R=\sin\theta \dd \theta\wedge \dd \phi$. The Lie 2-algebra of Hamiltonian 1-forms $\alpha=f_\alpha \gamma$ on $\FR^3\backslash\{0\}$ reduces accordingly to the Poisson algebra of functions on the 2-sphere.

\paragraph{Reduction of $S^3$.} Here let us choose the contact form $\gamma=\tfrac{1}{2}\dd\eta_3+\sin^2\eta_1\dd\eta_2$, so as to obtain on $S^2$ the symplectic structure $\varpi_R=\dd\gamma=2\sin\eta_1\cos\eta_1\dd\eta_1\wedge\dd\eta_2=\sin(2\eta_1)\dd\eta_1\wedge\dd\eta_2$. The Reeb vector here is $X_R=2\partial_{\eta_3}$, and for Hamiltonian 1-forms $\alpha$, $\beta$ along the M-theory direction we have
\begin{equation}
  \label{eq:10}
  \iota_{X_R}\pi_2(\alpha,\beta)=\frac{2}{\cos\eta_1\sin\eta_1}\eps^{ij3}\partial_{\eta_i}f_\alpha\partial_{\eta_j}f_\beta=\{f_\alpha,f_\beta\}~,
\end{equation}
which is the usual Poisson structure on $S^2$.

\subsection{Lie 2-algebras not originating from 2-plectic manifolds}\label{pp-wave 2-algebra}

Just as a Poisson manifold is not necessarily a symplectic manifold, we should not expect that any interesting Lie 2-algebra of 1-forms comes from a 2-plectic structure. To illustrate this point further, let us consider the categorification of Hpp-waves. 

Recall that ten-dimensional homogeneous plane waves arise as the
Penrose limit of the near horizon geometry $\mathrm{AdS}_5\times S^5$ in type IIB supergravity \cite{Blau:2002mw}. If we restrict the plane wave to four dimensions, it can be regarded as the group manifold of a twisted Heisenberg group. Its Lie algebra is the extension of the two-dimensional Heisenberg algebra by one additional generator $J$:
\begin{equation}\label{eq:Nappi_Witten_algebra}
 [\lambda_a,\lambda_b]=\eps_{ab}\unit~,~~~[J,\lambda_a]=\eps_{ab}\lambda_b~,~~~[\unit,\lambda_a]=[\unit,J]=0~.
\end{equation}
This algebra is also known as {\em Nappi-Witten algebra} and it can be regarded as linear Poisson structure on a four-dimensional Hpp-wave. Moreover, it can be obtained in various ways as a solution of the IKKT model, where $J$ and $\unit$ are regarded as quantized light-cone coordinates, while $\lambda_a$ are the quantized two remaining spatial coordinates. For further details, including an analogous twisted Nambu-Heisenberg algebra, see \cite{DeBellis:2010pf}.

A categorification of this Poisson structure on a four-dimensional
Hpp-wave would clearly correspond to a twist of the Lie 2-algebra
induced by the 2-plectic structure on $\FR^3$. Although the
integration theory of Lie 2-algebras is barely developed, one is led
to an interpretation of the twisted Lie 2-algebra as a categorified
linear Poisson structure on a five-dimensional Hpp-wave. We start from five coordinates $x^\pm$ and $x^i$, $i=1,\ldots,3$ together with the 1-forms
\begin{equation}
 \xi_i=\eps_{ijk}x^j\dd x^k~~~~\text{and} ~~~ \xi^i_\pm=x_\pm \dd x^i~.
\end{equation}
The twisted version of the Lie 2-algebra $\Pi_{\FR^3,\varpi}$ is given by
\begin{equation}\label{cat.d pp-wave algebra 1}
\pi_2(\xi_i,\xi_j)=-\eps_{ijk}\dd
x^k~,~~~~~\pi_3(\xi_i,\xi_j,\xi_k)=-\eps_{ijk}~,
\end{equation}
where we take the products involving the light-cone sector, parametrized by
$x^\pm$, to be:
\begin{equation}
\begin{aligned}\label{cat.d pp-wave algebra 2}
\pi_2(\xi_i,\xi_{j-})&=\eps_{ijk}\xi^k~, ~~~~~
\pi_2(\xi^i_-,\xi^j_-)=-\eps^{ij}_{\phantom{ij}k}\xi^k_-~, ~~~~~\pi_2(\xi^i_+,\dotsp)=0~,
\\ \\
\pi_3(\xi_i,\xi^j_-,\xi^k_-)&=0~,~~~~~
\pi_3(\xi^i_-,\xi_j,\xi_k)=\delta^i_k x^j-\delta^i_j x^k~,~~~~~\pi_3(\xi^i_-,\xi^j_-,\xi^k_-)=0~,
\end{aligned}
\end{equation}
while all the $\pi_3(\xi^i_+,\dotsp,\dotsp)=0$.
The two-products in the above reduce to the Nappi-Witten algebra in 4
dimensions \eqref{eq:Nappi_Witten_algebra} after contraction along one of the $\FR^3$ vectors, for
instance $\tfrac{\partial}{\partial x^3}$:
  \begin{equation}
    \xi_i\rightarrow \xi_a=\eps_{ab}x^b\dd x^3~,~~~\text{so that}~~~
    \lambda_a\equiv \iota_{\partial_3} \xi_a=\eps_{ab}x^b~,
  \end{equation}
if we further identify $J\equiv -x_-$. In analogy to the symplectic
case, we will set all $\pi_2(\xi^i_-,\dotsp)$ and $\pi_2(\xi^i,\dotsp)$ acting on exact 1-forms to
zero, in line with the interpretation that they should act as
derivations along the direction they define. By combining 2-products we
obtain expressions for $\pi_1(\pi_3(\dotsp,\dotsp,\dotsp))$ and thus
deduce 3-products $\pi_3$ that are compatible with the Lie 2-algebra
structure, given in the second line in \eqref{cat.d pp-wave
  algebra 2}. Note that these are only fixed up to constant terms by the
Lie 2-algebra equations, so here we chose the simplest possible form
for them. We can further take all mixed 2-products
$\pi_2(x^\pm,\xi_i)=\pi_2(\xi^i_\pm,x^j)=0$, as well as set
$\pi_2(\xi^i_\pm,x^\pm)=0$, since this does not affect the 2-algebra
equations, nor do we have any natural reason to expect them to be non-vanishing. 

Another example of a Lie 2-algebra that does not arise from a 3-form in the manner described in section \ref{ssec:2Plectic_Manifolds} is that of a twisted Poisson algebra \cite{Severa:2001qm} arising e.g.\ in the context of double geometry. This example points towards a more comprehensive mathematical description of higher Poisson structures. A Poisson structure on a manifold is encoded in a corresponding Poisson Lie algebroid. Analogously, one would expect that higher \mbox{(2-)Poisson} structures are encoded in a Courant Lie 2-algebroid. This is in fact the case for the twisted Poisson algebras discussed in \cite{Severa:2001qm}.

A geometric quantization of twisted Poisson manifolds has been proposed in \cite{Petalidou:0704.2989} and deformation quantization of these manifolds has been considered in \cite{Mylonas:2012pg}.

\subsection{Quantization}\label{ssec:2-plectic_quantization}

The quantization of 2-plectic manifolds remains an open
problem. Partial answers have been obtained by quantizing the
Nambu-Poisson bracket that arises from a 2-plectic structure under
certain conditions, cf.\ \cite{DeBellis:2010pf} and references
therein. Other approaches use a detour via loop spaces, see e.g.\ \cite{Saemann:2012ab}. For a more recent discussions of the general mechanism, see e.g.\ \cite{Nuiten:2013aa}. Attacking the quantization of 2-plectic manifolds directly
faces the aforementioned problem that even the algebraic structure of
classical observables is not fully clarified. Fortunately, we can ignore this problem and
regard classical quantization only as a Lie algebra homomorphism to
first order in $\hbar$ that maps the Poisson algebra to a Lie algebra
of quantum observables. The categorified analogue is then a Lie
2-algebra homomorphism to first order in $\hbar$ that maps a Lie 2-algebra of classical observables - arising e.g.\ from a 2-plectic structure - to a Lie 2-algebra of quantum observables. Roughly this point of view has been adopted e.g.\ in \cite{Rogers:2011zc}, see also \cite{Fiorenza:1304.6292}, where prequantization of 2-plectic manifolds has been developed to a considerable amount.
Usually, the symplectic form on certain quantizable manifolds defines the first Chern class of the prequantum line bundle. Fully analogously, a 2-plectic structure on certain manifolds defines the Dixmier-Douady class of a prequantum abelian gerbe. Many other ingredients of conventional geometric quantization have natural counterparts in this picture. In particular, the Atiyah algebroid, a symplectic Lie algebroid capturing the Souriau approach to geometric quantization, is replaced by a Courant Lie 2-algebroid, a symplectic Lie 2-algebroid.

Further evidence in favor of quantizing the Lie 2-algebra induced by the 2-plectic structure over the quantization of the Nambu-bracket stems from the above mentioned loop space approach. Both the 2-plectic structure as well as the prequantum abelian gerbe can be consistently mapped to a symplectic form of the loop space of the original manifold. Instead of quantizing the 2-plectic manifold, one can therefore quantize the induced symplectic loop space, cf.\ \cite{Saemann:2012ab,Saemann:2012ex} and references therein. This quantization of loop space is now naturally compatible with the quantization of the 2-plectic structure.

Having established that our notion of quantization will be necessarily incomplete, let us now specify it to the extend we can. Our guiding principle here will be a straightforward analogy with the correspondence principle \eqref{eq:correspondence_principle} of ordinary quantization: a quantization of a manifold $M$ endowed with a Lie 2-algebra $\Pi_{M}$ is a semistrict Lie 2-algebra $\hat{\Pi}_{M}$ with products $\mu_i$ together with a map 
\begin{equation}
 \hat{-}:\Pi_{M}\rightarrow \hat{\Pi}_{M}~,
\end{equation}
which is a Lie 2-algebra homomorphism to lowest order in a deformation parameter $\hbar$. For simplicity, we will restrict our attention to Lie 2-algebra homomorphisms $(\Psi_0,\Psi_{-1},\Psi_2)$ that are purely given in terms of chain maps with $\Psi_2=0$. This results in the following ``categorified correspondence principle:''
\begin{equation}\label{eq:2-correspondence_principle}
\begin{aligned}
\mu_1(\hat{X})=\widehat{-\di \hbar~\pi_1(X)}+\CO(\hbar)&~,~~~\mu_2(\hat{X},\hat{Y})=\widehat{-\di \hbar ~\pi_2(X,Y)}+\CO(\hbar^2)~,\\
\mu_3(\hat{X},\hat{Y},\hat{Z})&=\widehat{-\di \hbar ~\pi_3(X,Y,Z)}+\CO(\hbar^2)~.
\end{aligned}
\end{equation}
For our goals in this paper, this categorified correspondence principle will prove to be sufficient.

\subsection{Representation of the Heisenberg Lie 2-algebra}

While we cannot solve the problem of quantization of 2-plectic manifolds here, we can give some partial insight by regarding the analogue of the Heisenberg algebra, which arises in the quantization of $\FR^2$. More specifically, the Heisenberg algebra is spanned by quantized constant and linear functions, $\hat{x}^i$ and $\hat{c}=c\unit$, $c\in \FR$. These operators satisfy the commutation relation
\begin{equation}\label{eq:Heisenberg_algebra}
 [\hat{x}^a,\hat{x}^b]=\widehat{-\di \hbar \{x^a,x^b\}}=-\di \hbar \eps^{ab}\unit~,~~~a,b=1,2~.
\end{equation}
Note that the corrections to order $\CO(\hbar^2)$ in the correspondence principle vanish for coordinate functions. A representation for the Heisenberg algebra is given by $U_3$, the upper triangular $3\times 3$-dimensional matrices:
\begin{equation}\label{eq:rep-Lie}
 a \hat{x}^1+b \hat{x}^2-\di \hbar c\unit~~~\mapsto~~~\left(\begin{array}{ccc}0 & a & c \\ 0 & 0 & b \\ 0 & 0 & 0\end{array}\right)~,
\end{equation}
and the matrix commutator of these upper triangular matrices reproduces the algebra relation \eqref{eq:Heisenberg_algebra}.

The Heisenberg Lie 2-algebra is spanned by quantized constant and linear functions as well as constant and linear 1-forms $\hat{x}^i$, $\hat{c}=c\unit$ and $\xi^i$, $\dd x^i$, as defined in section \ref{ssec:2-plectic_examples}. The non-trivial Lie 2-algebra products for the quantized coordinate algebra are
\begin{equation}
 \mu_1(\hat{x}^i)=\widehat{-\di \hbar \dd x^i}~,~~~\mu_2(\hat{\xi}_i,\hat{\xi}_j)=-\widehat{\di \hbar\eps_{ijk} \dd x^k}~,~~~\mu_3(\hat{\xi}_i,\hat{\xi}_j,\hat{\xi}_k)=\widehat{\di \hbar\eps_{ijk}}\unit~,
\end{equation}
where we again assumed that the corrections in the correspondence principle to order $\CO(\hbar^2)$ vanish here.

We represent this Lie 2-algebra on the 2-vector space $\FR^4\xrightarrow{\mu_1} U_5$, where $\FR^4$ is spanned by basis vectors $e^0,e^i$ and $U_5$ is the vector space of upper triangular $5\times 5$-dimensional matrices. The chain maps of the Lie 2-algebra homomorphism are given by
\begin{equation}\label{eq:rep-2Lie}
\begin{aligned}
 -\di \hbar c\unit+ b_i \hat{x}^i~~~&\mapsto~~~c e^0+b_i e^i~,\\
 a^i \hat{\xi}_i-\di \hbar b_i \widehat{\dd x^i}~~~&\mapsto~~~\left(\begin{array}{ccccc}
								  0 & a^1 & b_3 & 0 & 0 \\ 
								  0 & 0 & a^2 & b_1 & 0 \\ 
								  0 & 0 & 0 & a^3  & b_2\\ 
								  0 & 0 & 0 & 0 & a^1\\
								  0 & 0 & 0 & 0 & 0
								  \end{array}\right)~.
\end{aligned}
\end{equation}
The non-trivial Lie 2-algebra products on this 2-vector space are
given by obvious maps $\mu_1:\FR^4\rightarrow U_5$ and
$\mu_3:U_5^{\wedge 3}\rightarrow \FR^4$ together with the map 
\begin{equation}
 \mu_2(u_1,u_2)=[P(u_1),P(u_2)]~,~~~ u_1,u_2\in U_5~,
\end{equation}
where
\begin{equation}
 P\left(\begin{array}{ccccc}
	0 & a^1 & b_3 & 0 & 0 \\ 
	0 & 0 & a^2 & b_1 & 0 \\ 
	0 & 0 & 0 & a^3  & b_2\\ 
	0 & 0 & 0 & 0 & a^1\\
	0 & 0 & 0 & 0 & 0
	\end{array}\right):=
	  \left(\begin{array}{ccccc}
	0 & a^1 & 0 & 0 & 0 \\ 
	0 & 0 & a^2 & 0 & 0 \\ 
	0 & 0 & 0 & a^3  & 0\\ 
	0 & 0 & 0 & 0 & a^1\\
	0 & 0 & 0 & 0 & 0
	\end{array}\right)~.
\end{equation}
Such brackets containing projectors are quite common in the context of derived brackets and strong homotopy Lie algebras, cf.\ \cite{Voronov:math0304038}. Note that the reduction of the representation \eqref{eq:rep-2Lie} to \eqref{eq:rep-Lie} is very transparent.

\section{Homogeneous Lie 2-algebra models}

Let us now come to the homogeneous Lie 2-algebra models, which are built from the various inner products. As stated before, these models are written in terms of a single type of field $X^a$, $a=1,\ldots,d$, which takes values in a Lie 2-algebra. We start by discussing the difference between the three kinds of metrics. We then consider the classical equations of motion and demonstrate that their solutions contain quantized symplectic and 2-plectic manifolds.

\subsection{Homogeneous Lie 2-algebra models and the various inner products}

The first ingredient are the various non-vanishing products on $L$, which we summarize here for the reader's convenience:
\begin{equation}
 \begin{aligned}
  \mu_1&:V\rightarrow W~,&\mu_1^*&:W\rightarrow V~,\\
  \mu_2&:V\wedge W \rightarrow V~,~~~&\mu_2&:W\wedge W\rightarrow W~,~~~&\mu_2^*&:V\wedge V\rightarrow W~,\\
  \mu_3&:W\wedge W\wedge W\rightarrow V~,~~~&\mu_3^*&:W\wedge W \otimes V\rightarrow W~.
 \end{aligned}
\end{equation}
Note that we can neglect the product $\kappa_2$, as it is built from the ones above. Moreover, note that $\mu^*_2(w,\ell)=-\mu_2(w,\ell)$ for any $w\in W$ and $\ell\in L$. The large number of remaining products makes it impossible to discuss a general action, and inspired by the M2-brane models, we will restrict ourselves to actions that are at most sextic in the fields.

In the following, we briefly discuss general Lie 2-algebra models that make use of the three inner products that we introduced in section \ref{ssec:inner_products}. Recall that a key feature of all inner products was the fact that
\begin{equation}\label{eq:W_invariance}
 \langle \mu_2(w,\ell_1),\ell_2 \rangle+\langle \ell_1,\mu_2(w,\ell_2)\rangle=0
\end{equation}
for $w\in W$ and $\ell_1,\ell_2\in L$. This property is required to guarantee that the actions of Lie 2-algebra models exhibit a nice symmetry algebra.

The cyclic metric $\langle \dotsp,\dotsp\rangle_\infty$ defined in section \ref{ssec:inner_products} is very restrictive. Recall that this metric corresponds to an invariant polynomial, which naturally induces actions for field theories of ``Chern-Simons type'', cf.\ \cite{Sati:0801.3480}. A typical example is the action discussed in \cite{Zwiebach:1992ie,Lada:1992wc}, whose stationary points are described by homotopy Maurer-Cartan equations. Here, however, we are more interested in actions of ``Yang-Mills type'', of which the IKKT model is an example.

Leaving out the product $\mu_1$, the only non-zero terms we can construct, up to fourth order in $X$, are
\begin{equation}\label{eq:action_cyclic}
  S_\infty\:=\ \tfrac{1}{2}m_{ab}\langle X^a,X^b\rangle_\infty + \tfrac{1}{3}c_{abc} \langle
  X^a,\mu_2(X^b,X^c)\rangle_\infty + \tfrac{1}{4}\langle \mu_2(X^a,X^b),\mu_2(X^a,X^b)\rangle_\infty~,
\end{equation}
where $m_{ab}$ is a `mass matrix' and $c_{abc}\in \FR$ is some totally antisymmetric tensor encoding a background yielding a cubic coupling. Higher order terms involving nested $\mu_2$ can be constructed, too. Note, however, that terms involving $\mu_3$ necessarily vanish, cf.\ \eqref{eq:too_strong}. Splitting the fields $X^a$ in the action \eqref{eq:action_cyclic} into the components $X^a=v^a+w^a$ with $v^a\in V$ and $w^a\in W$, we arrive at
\begin{equation}
\begin{aligned}
  S_\infty\:=\ &\tfrac{1}{2}m_{ab}\langle v^a,v^b\rangle_\infty+\tfrac{1}{2}m_{ab}\langle w^a,w^b\rangle_\infty + \tfrac{1}{3}c_{abc} \langle
  w^a,\mu_2(w^b,w^c)\rangle_\infty +\\
  &+\tfrac{1}{4}\langle \mu_2(w^a,w^b),\mu_2(w^a,w^b)\rangle_\infty~.
\end{aligned}
\end{equation}

In the case of the minimally invariant metric, we can write down more general terms. For example, we could consider the following action:
\begin{equation}\label{eq:action_W_invariant}
\begin{aligned}
   S_0\ =\ &\tfrac{1}{2}m_{ab}\langle X^a,X^b\rangle_0 + \tfrac{1}{3}c_{abc}\langle
  X^a,\mu_2(X^b,X^c))\rangle_0 + \tfrac{1}{4}\langle \mu_2(X^a,X^b),\mu_2(X^a,X^b)\rangle_0\\
&+d_{abcd}\langle X^a,\mu_3(X^b,X^c,X^d)\rangle_0 + \tfrac{1}{6}\lambda\langle \mu_3(X^a,X^b,X^c),\mu_3(X^a,X^b,X^c)\rangle_0\\
\ =\ &\tfrac{1}{2}m_{ab}\langle v^a,v^b\rangle_0+\tfrac{1}{2}m_{ab}\langle w^a,w^b\rangle_0+\tfrac{2}{3}c_{abc}\langle
  v^a,\mu_2(w^b,v^c))\rangle+\tfrac{1}{3}c_{abc}\langle
  w^a,\mu_2(w^b,w^c))\rangle\\
 &+\tfrac{1}{2}\langle \mu_2(w^a,v^b),\mu_2(w^a,v^b)\rangle+ \tfrac{1}{2}\langle \mu_2(w^a,v^b),\mu_2(v^a,w^b)\rangle+\tfrac14\langle \mu_2(w^a,w^b),\mu_2(w^a,w^b)\rangle\\
 &+\tfrac{1}{4}d_{abcd}\langle v^a,\mu_3(w^b,w^c,w^d)\rangle + \tfrac{1}{6}\lambda\langle \mu_3(w^a,w^b,w^c),\mu_3(w^a,w^b,w^c)\rangle~,
\end{aligned}
\end{equation}
where $c_{abc}\in\FR$ and $d_{abcd}\in\FR$ encode totally antisymmetric\footnote{While only totally antisymmetric parts of $c_{abc}$ contribute to $S_0$, this is not the case for $d_{abcd}$.} background tensors and $\lambda\in\FR$ is a coupling constant. 

In the case of the reduced metric, $V$ is considered as a sub vector space of $W$. Thus, we can replace $X^a$ in the action directly by $w^a$, and we get interaction terms like
\begin{equation}
d_{abcd}\langle X^a,\mu_3(X^b,X^c,X^d)\rangle_{\rm red}=d_{abcd}\langle w^a,\mu_3(w^b,w^c,w^d)\rangle_{\rm red}~,
\end{equation}
which, however, can be rewritten as $3d_{abcd}\langle w^a,\mu_2(\mu_2(w^{[b},w^c),w^{d]})\rangle$.

While actions built from minimally invariant and reduced inner products can contain considerably more interactions than those employing the cyclic inner product, it is not clear to us whether these additional terms are useful. In particular, when considering actions that have quantized symplectic and 2-plectic geometries as solutions, we can restrict ourselves to the terms contained in $S_\infty$.

\subsection{Symmetries of the models}\label{ssec:actions_and_symmetries}

The symmetries of a general Lie 2-algebra model have to be given by Lie 2-algebra automorphisms. Recall that the symmetry algebra relevant in the IKKT matrix model was the algebra of inner automorphisms of the underlying matrix algebra. We will therefore focus our attention here on {\em inner Lie 2-algebra automorphisms}, by which we mean automorphisms $\Psi:L\rightarrow L$ which read infinitesimally as
\begin{equation}
 \Psi_{-1}(v)=v+\mu_2(\alpha, v)~,~~\Psi_0(w)=w+\mu_2(\alpha,w)~~\mbox{and}~~ \Psi_2(w_1,w_2)=\mu_3(\alpha,w_1,w_2)~,
\end{equation}
where $v\in V$, $w\in W$, and $\alpha \in W$ is the (infinitesimal) gauge parameter. Under these symmetries, Lie 2-algebra actions remain invariant, independently of the inner product used in their definition. This is due to the invariance described in equation \eqref{eq:W_invariance}. For example, both the cyclic and minimally invariant inner products split into separate inner products of terms in $W$ and inner products of terms in $V$:
\begin{equation}
 S=\sum_i\langle w_{1,i},w_{2,i}\rangle + \sum_j\langle v_{1,j},v_{2,j} \rangle~.
\end{equation}
Each of these terms is invariant under inner Lie 2-algebra automorphisms, e.g.
\begin{equation}
 \delta \langle w_1,w_2\rangle=\langle \delta w_1, w_2\rangle + \langle w_1,\delta w_2 \rangle
 =\langle \mu_2(\alpha,w_1),w_2\rangle + \langle w_1,\mu_2(\alpha,w_2)\rangle=0~.
\end{equation}
One should stress in this context an important difference to conventional field theories: to propagate the action of the symmetry transformations from a higher product onto the fields, one has to take into account that a Lie 2-algebra automorphism also transforms the higher products themselves. For example, we have
\begin{equation}
 \delta\mu_2(w_1,w_2)=\mu_2(\delta w_1,w_2)+\mu_2(w_1,\delta w_2)+(\delta \mu_2)(w_1,w_2)~,
\end{equation}
and the explicit form of $(\delta\mu_2)(w_1,w_2)$ is easily read off equation \eqref{eq:Lie_2_algebra_homomorphism_conditions}. 

Recall that the IKKT model arose as a dimensional reduction of a
ten-dimensional supersymmetric gauge theory. Symmetries of this
model are therefore given by residual supersymmetry as well as
dimensionally reduced gauge symmetry. We might expect that something
similar happens in the case of Lie 2-algebra models, assuming that
they arise from a dimensional reduction of semistrict higher gauge
theory. While semistrict higher gauge theory has only been developed
partially, an attempt to capture its local gauge structure has been
made in \cite{Zucchini:2011aa}. 

In this framework, gauge symmetry is described by a Lie 2-algebra automorphism $(g_0,g_{-1},g_2)$ together with a flat connection doublet $(\sigma,\Sigma)$ and a 1-form $\tau$ taking values in $\sHom(W,V)$. The connection doublet and the 1-form are solutions of the consistency relations \eqref{eq:semistict_gauge_consistency}. For further reference, a concise overview over this gauge structure is included in appendix \ref{app:B}. 

After the dimensional reduction to a point, the consistency relations are satisfied for trivial $(\sigma,\Sigma)$ and $\tau$, and the whole gauge structure therefore reduces to a Lie 2-algebra automorphism. We thus arrive at the symmetries of our Lie 2-algebra model, in analogy with the case of the IKKT model. Note, however, that Lie 2-algebra models arising from dimensionally reducing a semistrict higher gauge theory to a point are more likely to be described by inhomogeneous Lie 2-algebra models, and we will return to this issue in section \ref{ssec:background_expansion_1}.

\subsection{Reduction to the IKKT model and quantized symplectic manifolds}

The reduction to the bosonic part of the IKKT model is a rather trivial affair. Given a (real) Lie algebra $\frg$, we can extend it trivially to a Lie 2-algebra $L_\frg:V\rightarrow W$ by putting $V=*=\{0\}$ and $W=\frg$. The only non-trivial higher product is then $\mu_2:\frg\times \frg\rightarrow \frg$, which is given by the commutator. The higher Jacobi identities are trivially satisfied. The Gram-Schmidt inner product yields an inner product on this Lie 2-algebra. This inner product satisfies simultaneously the axioms of cyclic, reduced and minimally invariant inner products, as one readily verifies. We can therefore work with any of the above discussed homogeneous Lie 2-algebra models.

All these models contained the following terms in the action:
\begin{equation}
  S_0\:=\ \tfrac{1}{2}m_{ab}\langle X^a,X^b\rangle + \tfrac{1}{3}c_{abc} \langle
  X^a,\mu_2(X^b,X^c)\rangle + \tfrac{1}{4}\langle \mu_2(X^a,X^b),\mu_2(X^a,X^b)\rangle~.
\end{equation}
Assuming that the underlying Lie 2-algebra is the Lie 2-algebra $L_\frg$, we recover the bosonic part of the IKKT matrix model \eqref{eq:action_IKKT} together with the bosonic part of the deformation terms \eqref{eq:action_IKKT_def}. Note that using a T-duality, one can then obtain BFSS matrix quantum mechanics \cite{Banks:1996vh} in the usual way.

We say that a solution to the IKKT model corresponds to a quantized
symplectic manifold, if the matrices $X^a$ describing this solution
are given by a complete set of quantized coordinate functions of a
noncommutative space. Note that for compact spaces like the fuzzy
sphere, these coordinate functions are given by embedding coordinates
of the compact manifold $M$ in some $\FR^n$. These coordinates should
be seen as the pullback of the coordinate functions on $\FR^n$ along
the embedding\footnote{According to the Whitney embedding theorem, any
  smooth manifold of dimension $d$ can be smoothly embedded in
  $\FR^{2d}$. This restricts the dimension of the quantized symplectic
  manifolds that can arise as solutions in the IKKT model. In fact,
  the Whitney embedding theorem can be improved to $\FR^{2d-1}$ unless
  $d$ is a power of $2$. } $e:M\embd \FR^n$.

Let us briefly recall three important solutions of the IKKT model for
future reference. For vanishing masses $m_{ab}$ and cubic couplings
$c_{abc}$, we obtain the Moyal plane $\FR^{2n}_\theta$, as already mentioned in the introduction. This space is described by quantized coordinate functions $\hat{x}^i$, $i=1,\ldots,2n$, satisfying the Heisenberg algebra, cf.\ \eqref{eq:Heisenberg_algebra}. 

The fuzzy sphere $S^2$ is described as a quantized submanifold of $\FR^3$ by the quantized coordinate algebra 
\begin{equation}
 [\hat{x}^i,\hat{x}^j]=-\di \hbar R \eps^{ijk}\hat{x}^k~,
\end{equation}
where $i,j,k=1,2,3$, $R$ is the radius of the fuzzy sphere and $\hbar=\frac{2}{k}$, $k\in \NN$, cf.\ \cite{DeBellis:2010pf}. As solutions to the IKKT model, it can be obtained in two ways. First of all, we can turn on a mass term
\begin{equation}
 m_{ij}=-2\hbar^2R^2\delta_{ij}~,
\end{equation}
as observed in \cite{Kimura:0103192}. Second, we can tune the cubic coupling proportional to the structure constants of $\asu(2)$, 
\begin{equation}
 c_{ijk}=-\di\hbar R\eps_{ijk}~,
\end{equation}
as discussed in \cite{Iso:0101102}. Both mass terms and cubic couplings can certainly be combined in a more general fashion.

The quantized Hpp-wave encoded in the Nappi-Witten algebra \eqref{eq:Nappi_Witten_algebra} is obtained as the solution
\begin{equation}
 \hat{x}^1=\lambda_1~,~~~\hat{x}^2=\lambda_2~,~~~\hat{x}^3=J\eand \hat{x}^4=\unit
\end{equation}
of the action $S_0$ with the following non-trivial mass-terms and couplings:
\begin{equation}\label{eq:background_IKKT_Hpp}
 m_{11}=m_{22}=-1\eand c_{ijk}=\eps_{ijk}~,~~~i,j,k=1,2,3~,
\end{equation}
see also \cite{DeBellis:2010sy}.

Before coming to the case of 2-plectic manifolds, let us briefly note
a subtle point. While the above quantized coordinate algebras do solve
the equations of motion resulting from the action $S_0$, they may not
correspond to quantized square integrable functions or may yield problematic terms in the action. For example, in the case of the Moyal plane, we have $[X^a,X^b]=\eps^{ab}\unit$. The term $\langle \mu_2(X^a,X^b),\mu_2(X^a,X^b)\rangle=\tr([X^a,X^b][X^a,X^b])$ is problematic when evaluated at this solution, as all non-trivial representations of the Heisenberg algebra are necessarily infinite-dimensional and the trace of $\unit$ is therefore ill-defined: the operator $\unit$ is not {\em trace class}. We will encounter the same issue in the case of Lie 2-algebra models. Recall, however, that we are not interested in the value of the action functional. We will first derive the equations of motion assuming our fields have finite norm and then continue the resulting equations to arbitrary Lie 2-algebra elements.

\subsection{Solutions corresponding to quantized 2-plectic manifolds}\label{ssec:ingredients:solutions}

As recalled above, we call a solution to the IKKT model a quantized symplectic manifold, if it is given in terms of quantized coordinate functions on $\FR^n$, into which the symplectic manifold is embedded. Similarly, solutions to the 3-Lie algebra model of \cite{DeBellis:2010sy} were given by quantized coordinate functions that took values in a 3-Lie algebra. Again, for compact spaces, quantized embedding coordinates of the manifold in some Euclidean space were used.

In the case of Lie 2-algebra models, the coordinate functions should be replaced by the quantization of certain elementary 1-forms. Let us characterize these 1-forms in the following. For compact spaces, we should again consider their embedding in some $\FR^n$ and use the pull-back of the elementary 1-forms on $\FR^n$ along the embedding. It therefore suffices to characterize elementary one-forms on $\FR^n$. There is a number of properties we would like these elementary 1-forms to have:
\begin{conditions}
 \item[(i)] They should be as simple as possible.
 \item[(ii)] They cannot be exact, as exact forms are central in the Lie 2-algebras induced by 2-plectic structures.
 \item[(iii)] Just as with Cartesian coordinate functions on $\FR^n$, the Hamiltonian vector field of the 1-forms should equal the derivative with respect to the Cartesian coordinates on $\FR^n$.
 \item[(iv)] Under the reduction procedure outlined in section \ref{ssec:reduction}, they should reduce to coordinate functions on $\FR^{n-1}$. 
\end{conditions}

In Cartesian coordinates $x^i$ on $\FR^n$, the simplest 1-forms on $\FR^n$ that are not exact are given by
\begin{equation}\label{eq:elementary_1-forms}
 \xi^{ij}=x^{[i}\dd x^{j]}~,
\end{equation}
and we have encountered these already in section \ref{ssec:2-plectic_examples}. One can easily verify that (iv) on its own would also lead to \eqref{eq:elementary_1-forms}. Moreover, on spaces $\FR^{3n}$ with canonical 2-plectic structure, these elementary 1-forms satisfy (iii). 

Another requirement one might impose is on the quantization of elementary 1-forms: the correspondence principle \eqref{eq:2-correspondence_principle} should hold exactly and should not receive any corrections to order $\CO(\hbar^2)$.

Note that the Lie 2-algebras we obtain from a 2-plectic structure are not reduced, and we do not expect that the corresponding quantized Lie 2-algebra will be reduced. In discussing solutions, we therefore have to restrict ourselves to the cyclic and minimally invariant inner products. In both cases, we are interested in the same action\footnote{We were not able to use the additional terms in the action \eqref{eq:action_W_invariant} in any sensible way to accommodate the desired solutions of quantized geometries; neither did they seem necessary.},
\begin{equation}
  S_1\:=\ \tfrac{1}{2}m_{ab}\langle X^a,X^b\rangle + \tfrac{1}{3}c_{abc} \langle
  X^a,\mu_2(X^b,X^c)\rangle + \tfrac{1}{4}\langle \mu_2(X^a,X^b),\mu_2(X^a,X^b)\rangle~,
\end{equation}
which, however, leads to different equations of motion. In the cyclic case, we have 
\begin{equation}\label{eq:eom_cyclic}
  m_{ab}w^b+\mu_2(w^b,\mu_2(w^b,w^a))+c_{abc}\mu_2(w^b,w^c)=0\eand m_{ab}v^b=0~,
\end{equation}
while in the minimally invariant case, we have
\begin{equation}\label{eq:eom_0}
\begin{aligned}
m_{ab}v^b+\tfrac{4}{3}c_{abc}\mu_2(w^b,v^c)+\tfrac12\mu_2(w^b,\mu_2(w^b,v^a)) +\tfrac12\mu_2(w^b,\mu_2(v^b,w^a))&=0~,\\
m_{ab}w^b-\tfrac{2}{3}c_{abc}\mu_2^*(v^c,v^b)+c_{abc}\mu_2(w^b,w^c)+\tfrac12\mu_2^*(v^b,\mu_2(v^b,w^a))
+\mu_2(w^b,\mu_2(w^b,w^a))&=0~.
\end{aligned}
\end{equation}
We now restrict to Lie 2-algebras that arise from the quantization of a Lie 2-algebra $\Pi_{M,\varpi}$ and impose the above mentioned requirement that for elementary functions and 1-forms, the correspondence principle \eqref{eq:2-correspondence_principle} holds precisely without corrections to order $\CO(\hbar^2)$. This implies that in equations \eqref{eq:eom_0}, the terms containing the products
\begin{equation}
 \mu_2:W\times V\rightarrow V\eand \mu_2^*:V\times V\rightarrow W
\end{equation}
vanish on elementary 1-forms and equations \eqref{eq:eom_0} reduce to \eqref{eq:eom_cyclic}. We can therefore restrict our attention to the latter equations of motion.

\subsection{Examples of quantized categorified Poisson manifolds as solutions}

As a first example, we consider the quantization of $\Pi_{\FR^3,\varpi}$, where $\varpi$ is again the canonical volume form on $\FR^3$. Just as the Moyal plane was obtained from the undeformed IKKT model, we expect the quantization $\hat{\Pi}_{\FR^3,\varpi}$ of $\Pi_{\FR^3,\varpi}$ to arise from the action $S_1$ with $m=c=0$. This is indeed the case: the quantization of the 1-forms $\xi_i=\tfrac{1}{2}\eps_{ijk}x^i\dd x^k$ satisfy the following algebra
\begin{equation}
 \mu_2(\hat{\xi}_i,\hat{\xi}_j)=-\di\hbar\eps_{ijk}\widehat{\dd x^k}~,
\end{equation}
where $\widehat{\dd x^k}$ is central in $\hat{\Pi}_{\FR^3,\varpi}$. Putting
\begin{equation}
 w^i=\hat{\xi}_i\eand v^i=0~,~~~i=1,\ldots,3~,
\end{equation}
we obtain a solution to \eqref{eq:eom_cyclic}, which we interpret as a quantization $\FR^3_\hbar$ of $\FR^3$ as 2-plectic manifold. 

Note that the solution of the IKKT model corresponding to the Moyal plane trivially extends to Cartesian products $\FR^{2n}_\theta=\FR^2_\theta\times\cdots\times \FR^2_\theta$. The same holds here, and we obtain quantized 2-plectic manifolds $\FR^{3n}_\hbar=\FR^3_\hbar\times \cdots\times \FR^3_\hbar$.

Note also that as a special case to the above solution, we can use the subalgebra of $\hat{\Pi}_{\FR^3,\varpi}$, which corresponds to the reduction to the fuzzy sphere as discussed in section \ref{ssec:reduction}. This yields a continuous foliation of quantized $\FR^3_\hbar$ by fuzzy spheres, which is different to the discrete foliation given by the space $\FR^3_\lambda$ as introduced in \cite{Hammou:2001cc}.

As our second example, let us consider the quantization of the 2-plectic sphere $S^3$. First, note that analogously to the case of the fuzzy sphere solution to the IKKT model, we should embed the 3-sphere into $\FR^4$ and describe its quantization as a push-forward on elementary 1-forms on $\FR^4$. More specifically, we consider the 1-forms
\begin{equation}\label{eq:1-forms_S^3}
 \xi_{\mu\nu}:=\tfrac{1}{2}\eps_{\mu\nu\kappa\lambda}x^\kappa\dd x^\lambda~,
\end{equation}
where $x^\mu,~\mu=1,\ldots,4~,$ are the embedding coordinates of $e:S^3\hookrightarrow \FR^4$, where $e(S^3)=\{||x||=1\,|\,x\in \FR^4\}$. The higher product $\pi_2$ on these elementary 1-forms is given by
\begin{equation}\label{eq:cat_of_so4}
 \pi_2(\xi_{\mu\nu},\xi_{\kappa\lambda})=\delta_{\nu\kappa}\xi_{\mu\lambda}-\delta_{\mu\kappa}\xi_{\nu\lambda}-\delta_{\nu\lambda}\xi_{\mu\kappa}+\delta_{\mu\lambda}\xi_{\nu\kappa}+\pi_1(R_{\mu\nu\kappa\lambda})~,
\end{equation}
where
\begin{equation}
 R_{\mu\nu\kappa\lambda}=\tfrac{1}{4}\left(\eps_{\nu\kappa\lambda\rho}x^\rho x^\mu-\eps_{\mu\kappa\lambda\rho}x^\rho x^\nu-\eps_{\kappa\mu\nu\rho}x^\rho x^\lambda+\eps_{\lambda\mu\nu\rho}x^\rho x^\kappa\right)~.
\end{equation}
Equation \eqref{eq:cat_of_so4} shows that this Lie 2-algebra of elementary 1-forms is in fact a categorification of the Lie algebra $\aso(4)$, where the usual commutation relations hold up to the isomorphism $\pi_1(R_{\mu\nu\kappa\lambda})$. 

Comparing again with the case of the fuzzy sphere arising in the IKKT model, we expect that the quantized 3-sphere arises in two different ways. First, a solution to $S_1$ is given in terms of the quantized 1-forms defined in \eqref{eq:1-forms_S^3} by
\begin{equation}\label{eq:solution_S^3}
 (w^I)=(\hat{\xi}_{12},\hat{\xi}_{13},\hat{\xi}_{14},\hat{\xi}_{23},\hat{\xi}_{24},\hat{\xi}_{34})\eand v^I=0~,
\end{equation}
if we set the masses to 
\begin{equation}
 m_{IJ}=-4\hbar^2\delta_{IJ}~,~~~I,J=1,\ldots,6~.
\end{equation}
Note that the index $I$ should here be regarded as a multi-index $I=([mn])$. Second, \eqref{eq:solution_S^3} is also a solution, if we tune the $c_{IJK}$ to the structure constants of $\aso(4)$ in the representation categorified in \eqref{eq:cat_of_so4}. Explicitly, we have the following non-trivial entries:
\begin{equation}
 c_{[124]}=\di\hbar~,~~~c_{[135]}=\di\hbar~,~~~c_{[236]}=\di\hbar~,~~~c_{[456]}=\di\hbar~.
\end{equation}
It is quite striking that quantized $S^3$ arises in the same manner in our Lie 2-algebra models as the fuzzy sphere arose in the IKKT model.

As our last example, let us consider the Lie 2-algebra corresponding to a categorification of the Nappi-Witten algebra, which we interpreted as the Lie 2-algebra related to a five-dimensional Hpp-wave. We have nine elementary 1-forms,
\begin{equation}\label{eq:sol_Hpp}
 (w^m)=(\hat{\xi}_1,\hat{\xi}_2,\hat{\xi}_3,\hat{\xi}_+^1,\hat{\xi}_+^2,\hat{\xi}_+^3,\hat{\xi}_-^1,\hat{\xi}_-^2,\hat{\xi}_-^3)~,
\end{equation}
and their non-trivial products $\mu_2$ read as
\begin{equation}
 \mu_2(\hat{\xi}_i,\hat{\xi}_j)=\di\hbar\eps_{ijk}\dd x^k~,~~~\mu_2(\hat{\xi}_i,\hat{\xi}_j^-)=-\di \hbar \eps_{ijk}\hat{\xi}^k\eand \mu_2(\hat{\xi}^i_-,\hat{\xi}_-^j)=\di\hbar\eps^{ij}{}_k\hat{\xi}^k_-~.
\end{equation}
The most general action $S_1$ to which \eqref{eq:sol_Hpp} is a solution has the following mass parameters and cubic coupling terms:
\begin{equation}
\begin{aligned}
 m_{mn}=\diag(4\hbar^2,4\hbar^2,4\hbar^2,0,0,0,-2\di \hbar\,c_{789},-2\di \hbar\,c_{789},-2\di \hbar\,c_{789})~,&\\
 c_{[ijk_-]}=-\di\hbar\eps_{ijk_-}~,~~~c_{[ijk_+]}=c_{[ijk]}=c_{[ij_-k_-]}=c_{[i_+j_-k_-]}=0~,\hspace{0.1cm}
\end{aligned}
\end{equation}
while the remaining cubic couplings can be chosen arbitrarily. Here, indices $i_+$ run over $4,5,6$ and indices $i_-$ run over $7,8,9$. Note that these background fields are very similar to those in \eqref{eq:background_IKKT_Hpp} that gave rise to the Hpp-wave solution in the IKKT model.

\section{Inhomogeneous Lie 2-algebra models}\label{Inhomogeneous models}

We now come to inhomogeneous Lie 2-algebra models, in which we have two kinds of fields $\{X^a\}$ and $\{Y^i\}$ taking values in $V$ and $W$, respectively. This class of models includes the homogeneous models as those actions that are written in terms of sums $X^a+Y^a$. Therefore the inhomogeneous models can exhibit all the solutions we found in the previous section. We will start with an inhomogeneous Lie 2-algebra model that reduces for skeletal and strict Lie 2-algebras to zero-dimensional M2-brane models. We then consider a specific inhomogeneous Lie 2-algebra model that results from dimensionally reducing a higher gauge theory and analyze fluctuations around a special solution.

Note that inhomogeneous Lie 2-algebra models are also invariant under the inner Lie 2-algebra automorphisms discussed in section \ref{ssec:actions_and_symmetries}.

\subsection{Dimensionally reduced M2-brane models}

We showed in section \ref{ssec:M2-brane_3_algebras} that Lie 2-algebras that are either skeletal or strict come with a real 3-algebra structure, where the ternary bracket is given by
\begin{equation}\label{eq:ternary_bracket2}
 [v_1,v_2,v_3]=-\mu_2(\mu_2^*(v_1,v_2),v_3)~.
\end{equation}
For $\mu_2^*$ to be non-trivial, we will have to work with the minimally invariant metric $\langle -,-\rangle_0$. 

We can now write down inhomogeneous Lie 2-algebra models that make use of this ternary bracket and reduce to previously studied zero-dimensional models related to M2-brane models. The action we are interested in reads as
\begin{equation}\label{eq:M2-brane_model}
\begin{aligned}
 S_{\rm M2}=&\tfrac{1}{6}\eps^{ijk}\langle Y^i,\mu_2(Y^j,Y^k)\rangle_0-\tfrac{1}{2}\langle\mu_2(Y^i,X^a),\mu_2(Y^i,X^a)\rangle_0+\tfrac{\di}{2}\langle \bar{\Psi},\mu_2(\Gamma^i Y^i,\Psi)\rangle_0\\&
 -\tfrac{\di}{4}\langle \bar{\Psi},\mu_2(\mu_2^*(X^a,X^b),\Gamma_{ab}\Psi)\rangle_0-\tfrac{1}{12}\langle \mu_2(\mu_2^*(X^a,X^b),X^c),\mu_2(\mu_2^*(X^a,X^b),X^c)\rangle_0~,
\end{aligned}
\end{equation}
where the scalars $X^a$, $a=1,\ldots,8$, and the spinors $\Psi$ take values in $V$, while the scalars $Y^i$, $i=0,\ldots,2$, take values in $W$. Our spinor and Clifford algebra conventions are those of \cite{Bagger:2007jr}.

For skeletal or strict Lie 2-algebras, the action \eqref{eq:M2-brane_model} equals that of a full dimensional reduction of the $\CN=2$ M2-brane models discussed in \cite{Cherkis:2008qr}. Our action then inherits $\CN=2$ supersymmetry from the 3-dimensional model.

If the ternary bracket \eqref{eq:ternary_bracket2} happens to be antisymmetric\footnote{which is the case e.g.\ if the underlying real 3-algebra is $A_4$}, we recover the 3-Lie algebra models that we discussed in the introduction. In particular, the models of \cite{Lee:2009ue,Furuuchi:2009ax,Sato:2009tr} are obtained by putting $Y_i=0$ and letting $a=0,\ldots,10$, otherwise one arrives at the model discussed in \cite{DeBellis:2010sy}. It is a trivial exercise to add deformation terms to \eqref{eq:M2-brane_model} that are written in terms of 3-brackets.

Note that inhomogeneous Lie 2-algebra models reproducing the dimensionally fully reduced ABJM model can also be written down in a straightforward fashion.

To obtain interesting solutions to the model \eqref{eq:M2-brane_model} one should either consider solutions with $Y_i\neq 0$ or solutions that do not arise from 2-plectic manifolds. Otherwise, the equations of motion are trivially satisfied. It is not clear how to interpret solutions that contain both nontrivial $Y_i$ and $X^a$. We therefore refrain from going into any more detail at this point.

\subsection{Background expansion for higher gauge theory}\label{ssec:background_expansion_1}

Recall from section \ref{ssec:background_IKKT} that expanding the action of the IKKT model around a solution corresponding to a noncommutative space yields essentially the action for Yang-Mills theory on that noncommutative space \cite{Aoki:1999vr}. In particular, consider the action \eqref{eq:action_IKKT}. A solution to this action is the Moyal space $\FR^{2n}_\theta$ with coordinates satisfying $[\hat{x}^\mu,\hat{x}^\nu]=-\di\theta^{\mu\nu}$, $\mu,\nu=1,\ldots,2n$. If we now expand around this solution by writing $X^\mu=\hat{x}^\mu+\hat{A}^\mu$ and observing that $\theta_{\mu\nu}[\hat{x}^\nu,\hat{f}]=\widehat{\dpar_\mu f}$, we obtain noncommutative Yang-Mills theory on $\FR^{2n}_\theta$.

An attempt has been made to reproduce this observation in the context of quantized Nambu-Poisson manifolds using 3-Lie algebras in \cite{DeBellis:2010sy}, but the construction seemed far less natural than in the IKKT case.

Let us now try to obtain field theories on quantized 2-plectic spaces by performing a background expansion. For this, we have to choose the kind of action we expect to reproduce. The most natural candidate here are higher BF-theories as discussed in \cite{Martins:2010ry}. 

For simplicity, we will consider higher BF theory on $\FR^3$. The field content consists of a 1-form $A$ and a 2-form $B$. Usually, these take values in the vector spaces $W$ and $V$, respectively, that form a strict Lie 2-algebra. Here, however, we immediately allow for a semistrict Lie 2-algebra, and neglect all the technical difficulties that come with a complete discussion of semistrict higher gauge theory, see \cite{Zucchini:2011aa} and appendix \ref{app:B}. If the higher BF theory is supposed to describe a connective structure that captures the parallel transport of an extended object, we have to impose the fake curvature condition
\begin{equation}
 0=\CF:=F-\mu_1(B):=\dd A+\mu_2(A,A)-\mu_1(B)~.
\end{equation}
As usual in BF-theory, we also expect the 3-form curvature to vanish:
\begin{equation}
 0=H:=\dd B+\mu_2(A,B)+\tfrac16\mu_3(A,A,A)~,
\end{equation}
where we extended the usual definition of the 3-form curvature $H$ in higher gauge theory by a term $\mu_3(A,A,A)$, cf. \cite{Zucchini:2011aa}. Altogether, we arrive at the action
\begin{equation}
 S_{\rm BF}=\int_{\FR^3} \langle \lambda_1,\CF\rangle_0+\langle \lambda_0,H\rangle_0~,
\end{equation}
where $\lambda_0$ and $\lambda_1$ are 0- and 1-forms taking values in $V$ and $W$, respectively.

To obtain a Lie 2-algebra model, we dimensionally reduce the action $S_{\rm BF}$ to a point. We are left with fields $X_{ij}$, $i,j=1,\ldots,3$, taking values in $V$ and fields $Y_i$ with values in $W$ together with additional Lagrange multiplier fields $\lambda_i$ and $\lambda$. Note that we should also twist the BF action by terms $2\di\hbar\langle \lambda_i,\widehat{\dd x^i}\rangle_0$ and $\di\hbar\langle \lambda, 1\rangle_0$. This can be easily seen by imagining writing down a BF theory on the Moyal plane; even in the Yang-Mills case we introduce such a twist, cf.\ \eqref{eq:action_IKKT}. The total action reads as
\begin{equation}\label{eq:BF_red}
\begin{aligned}
 S_{0d}=&\eps^{ijk}\langle \lambda_i,\mu_2(Y_j,Y_k)-\tfrac12\mu_1(X_{jk})\rangle_0+2\di\hbar\langle \lambda_i,\widehat{\dd x^i}\rangle_0\\ &+\eps^{ijk}\langle \lambda,\tfrac12\mu_2(Y_i,X_{jk})+\tfrac16\mu_3(Y_i,Y_j,Y_k)\rangle_0-\di\hbar\langle \lambda,\unit\rangle_0~.
\end{aligned}
\end{equation}
A solution to the corresponding equations of motion is given by elements of the semistrict Lie 2-algebra $\hat{\Pi}_{\FR^3,\varpi}$ arising from the quantization of the Lie 2-algebra $\Pi_{\FR^3,\varpi}$:
\begin{equation}\label{eq:sol_BF1}
 Y_i=\hat{\xi}_i~,~~~X_{ij}=0\eand \lambda_i=\lambda=0~.
\end{equation}
We now observe that
\begin{equation}
 \dd x^i\wedge\pi_2(\xi_i,\alpha)=\dd \alpha~,
\end{equation}
and therefore $\mu_2(\widehat{\xi_i},\dotsp)$ should be identified with a quantum derivation, at least on 1-forms. Consider the background field expansion
\begin{equation}
 Y_i=\hat{\xi}_i+\hat{A}_i~,~~~X_{ij}=0+\hat{B}_{ij}~,
\end{equation}
where $\hat{A}_i$ and $\hat{B}_{ij}$ take values in the obvious vector spaces contained in $\hat{\Pi}_{\FR^3,\varpi}$. The action \eqref{eq:BF_red} becomes
\begin{equation}\label{eq:2LBF_action}
 S_{\rm 2LBF}=\eps^{ijk}\langle \lambda_i,\hat{\CF}_{jk}\rangle_0 + \eps^{ijk}\langle \lambda,\hat{H}_{ijk}\rangle_0~,
\end{equation}
where we defined
\begin{equation}
 \begin{aligned}
 \hat{\CF}_{ij}&=\mu_2(\hat{\xi}_i,\hat{A}_j)-\mu_2(\hat{\xi}_j,\hat{A}_i)+\mu_2(\hat{A}_i,\hat{A}_j)-\mu_1(\hat{B}_{ij})~,\\
 \hat{H}_{ijk}&=\tfrac12\mu_2(\hat{\xi}_{[i},\hat{B}_{jk]})+\tfrac12\mu_2(\hat{A}_{[i},\hat{B}_{jk]})+\tfrac16\mu_3(\hat{\xi}_i+\hat{A}_i,\hat{\xi}_j+\hat{A}_j,\hat{\xi}_k+\hat{A}_k)-\di\hbar\hat{\unit}~.  
 \end{aligned}
\end{equation}
It is interesting to note how the Lie 2-algebra $\Pi_{\FR^3,\varpi}$ turned into a gauge Lie 2-algebra of higher BF-theory on a quantized 2-plectic space.

Recall that in the Lie 2-algebra arising from 2-plectic $\FR^3$, the higher product between functions and 1-forms, $\pi_2:\Omega^1(\FR^3)\times \CC^\infty(\FR^3)\rightarrow \CC^\infty(\FR^3)$, is trivial. We therefore have in the quantized case
\begin{equation}
 \hat{H}_{ijk}=\tfrac16\mu_3(\hat{\xi}_i+\hat{A}_i,\hat{\xi}_j+\hat{A}_j,\hat{\xi}_k+\hat{A}_k)+\CO(\hbar^2)~.  
\end{equation}
This interpretation is very close to the one used in \cite{Chu:2011yd}. There, the 3-form curvature $H$ was identified in a 3-Lie algebra valued model with the product $[\hat{A}_i,\hat{A}_j,\hat{A}_k]$. Considering the 3-Lie algebra $2\sMat(n)$ constructed in section \ref{ssec:M2-brane_3_algebras}, where the triple bracket of the 3-Lie algebra can be identified with the higher product $\mu_3$, our $\hat{H}$ essentially matches that of \cite{Chu:2011yd}.

\subsection{Background expansion using an isomorphic Lie 2-algebra structure}

As we saw in the previous section, the Lie 2-algebra $\CC^\infty(M)\rightarrow \frH(M)$ that we obtained from a 2-plectic manifold $M$ in section \ref{ssec:2Plectic_Manifolds} is very restrictive. In particular, elements of $V$ have no possibility of interacting via higher products with $W$. To remedy this, note that we can add the non-trivial product
\begin{equation}\label{eq:mod_product_1}
 \pi_2(f,\alpha):=-\iota_{X_\alpha}\dd f~,~~~f\in \CC^\infty(M)~,~\alpha\in \frH(M)~.
\end{equation}
This additional product, however, violates the higher homotopy relation $\pi_1(\pi_2(\alpha,f))=\pi_2(\alpha,\pi_1(f))$. To fix this, we modify $\pi_2:\frH(M)\times \frH(M)\rightarrow \frH(M)$ as follows:
\begin{equation}\label{eq:mod_product_2}
 \pi_2(\alpha,\beta):=-\iota_{X_\alpha}\iota_{X_\beta}\varpi-\dd(\iota_{X_\alpha}\beta)+\dd(\iota_{X_\beta}\alpha)~,~~~\alpha,\beta\in\frH(M)~,
\end{equation}
where we note that $\pi_2(\alpha,\beta)$ is indeed in $\frH(M)$. The products $\pi_1$ and $\pi_3$ remain unmodified:
\begin{equation}
 \pi_1(f):=\dd f\eand\pi_3(\alpha,\beta,\gamma):=-\iota_{X_\alpha}\iota_{X_\beta}\iota_{X_\gamma}\varpi
\end{equation}
for $f\in \CC^\infty(M)$ and $\alpha,\beta,\gamma\in\Omega^1(M)$. We will denote the resulting structure by $\tilde{\Pi}_{M,\varpi}$. 

Instead of verifying all the higher homotopy relations \eqref{eq:homotopy_relations} for $\tilde{\Pi}_{M,\varpi}$, we can prove a stronger statement: $\tilde{\Pi}_{M,\varpi}$ is isomorphic to the Lie 2-algebras $\Pi_{M,\varpi}$. This is easily seen by giving the explicit Lie 2-algebra homomorphism, cf.\ section \ref{ssec:Lie_2_algebra_homomorphism}:
\begin{equation}
 \Psi_{-1}=\id~,~~~\Psi_0=\id~,~~~\Psi_2(\alpha,\beta)=\iota_{X_\alpha} \beta-\iota_{X_\beta} \alpha~.
\end{equation}
Equations \eqref{eq:Lie_2_algebra_homomorphism_conditions} then yield the higher products \eqref{eq:mod_product_1} and \eqref{eq:mod_product_2}. The higher product $\mu_3$ remains unmodified, as one readily verifies by direct computation using the identity $\iota_{[X_\alpha,X_\beta]}=\CL_{X_\alpha}\iota_{X_\beta}-\iota_{X_\beta}\CL_{X_\alpha}$.

As an example, let us briefly study the Lie 2-algebra $\tilde{\Pi}_{\FR^3,\varpi}$ with 2-plectic form $\varpi=\dd x^1\wedge \dd x^2\wedge \dd x^3$. The Hamiltonian vector fields as well as $\pi_1$ and $\pi_3$ are listed in section \ref{ssec:2-plectic_examples}. The formulas for the new products read as
\begin{equation}
\begin{aligned}
 \pi_2(f,\alpha)&=-\eps^{ijk}\dpar_i f \dpar_j\alpha_k~,\\
  \pi_2(\alpha,\beta)&=\eps^{ijk}\left(\dpar_i\alpha_k(\dpar_j\beta_\ell-\dpar_\ell\beta_j)+\dpar_\ell(\alpha_i\dpar_j\beta_k-\beta_i\dpar_j\alpha_k)\right)\dd x^\ell~.
\end{aligned}
\end{equation}
For the constant and linear 1-forms $\dd x^i$ and $\xi_i=\tfrac{1}{2}\eps_{ijk}x^j\dd x^k$ we have
\begin{equation}
 \pi_2(f,\dd x^i)=0~,~~~\pi_2(f,\xi_i)=-\dpar_i f~,~~~\pi_2(\dd x^i,\alpha)=\eps^{ijk}\dpar_\ell \dpar_j x_k \dd x^\ell~,~~~\pi_2(\xi_i,\xi_j)=0~.
\end{equation}
In particular, we see that the operator $\pi_2(\xi_i,f)$ can here be interpreted as a derivation on functions, while it lost its nice derivation property property on 1-forms. 

To define a BF-theory via a background expansion using the Lie 2-algebra $\tilde{\Pi}_{\FR^3,\varpi}$, we should consider the action \eqref{eq:BF_red} without the twist term $2\di\hbar\langle \lambda_i,\widehat{\dd x^i}\rangle_0$. A classical configuration of this action is again \eqref{eq:sol_BF1} and we can follow the discussion of the previous section. 

\section{Conclusions}

In this paper, we initiated a study of zero-dimensional field theories, in which the fields take values in a semistrict Lie 2-algebra, or, equivalently, a 2-term $L_\infty$-algebra. These Lie 2-algebra models are a categorification of the IKKT matrix model, which is conjectured to provide a background independent formulation of string theory. In particular, Lie 2-algebra models contain the (bosonic part of the) IKKT model and all of its bosonic deformations.

We explored the various notions of inner products on Lie 2-algebras as well as the resulting structure of transposed products. Here, we made an observation concerning the connection between 3-algebras appearing in M2-brane models and categorified Lie algebras. Besides the established link between Lie 2-algebras and the 3-algebras of M2-brane models via differential crossed modules, we also showed that any skeletal Lie 2-algebra with inner product comes with a 3-algebra structure. Moreover, there is a class of reduced Lie 2-algebras, in which the higher product $\mu_3$ can be identified with the 3-bracket of a 3-Lie algebra.

We also pointed out the interaction of inner Lie 2-algebra homomorphisms with the various inner products. This allowed us to examine the symmetries of Lie 2-algebra models, which are compatible with those expected from a dimensional reduction of semistrict higher gauge theory. This is to be compared with the IKKT model, where the symmetry algebra arises from a dimensional reduction of the gauge theory.

We divided the Lie 2-algebra models into two classes: homogeneous Lie 2-algebra models are defined in terms of a single class of fields that take values in the Lie 2-algebra. Inhomogeneous Lie 2-algebra models feature two types of fields, each living within one of the graded vector subspaces of the 2-term $L_\infty$-algebra underlying the model.

Just as in the case of the IKKT model, where solutions to the classical equations of motion can be identified with quantized symplectic manifolds, the homogeneous Lie 2-algebra models we studied have solutions that can be interpreted as quantized 2-plectic manifolds. While the quantization of 2-plectic manifolds is still not fully understood, it was straightforward to outline the expected features of such a quantization that are required for our purposes. In particular, it is expected that under quantization the Lie 2-algebra induced by the 2-plectic structure on a manifold is mapped to a Lie 2-algebra of quantum observables with this map being a Lie 2-algebra homomorphism to lowest order in a deformation parameter $\hbar$. As an example, we examined the Heisenberg Lie 2-algebra, which is contained in the Lie 2-algebra arising from the quantization of $\FR^3$. We gave a representation in terms of derived brackets on a 2-vector space. 

The quantized symplectic manifolds most readily obtained as solutions in the IKKT model are the Moyal plane and the fuzzy sphere, as well as their Cartesian products. In the Lie 2-algebra models, we found solutions that correspond to the quantizations of $\FR^3$ and $S^3$, where the 2-plectic form was given by the canonical volume form on these spaces. Remarkably, these solutions appeared in complete analogy with the above mentioned solutions of the IKKT model.

We also studied solutions given by Lie 2-algebras that do not arise from 2-plectic manifolds. In particular, we considered the Nappi-Witten algebra, which can be regarded as linear Poisson structure on a four-dimensional Hpp-wave. This algebra gives a solution of a particular deformation of the IKKT model. We constructed a categorified analogue  of the Nappi-Witten algebra corresponding to a five-dimensional Hpp-wave and again we found that it appears as a solution of our Lie 2-algebra models.

We were able to show that certain inhomogeneous Lie 2-algebra models reproduce previously considered zero-dimensional field theories that are related to M2-brane models. Furthermore, we considered Lie 2-algebra models that arise from a dimensional reduction of a semistrict higher BF-theory in three dimensions. These models contained again a quantization of $\FR^3$ as a classical configuration, and expanding around this configuration, we obtained an action that can be interpreted as semistrict higher BF-theory on the quantized $\FR^3$. This is fully analogous to the case of the IKKT model, where it is known that expanding around a solution corresponding to a quantized symplectic manifold yields the action of Yang-Mills theory on this noncommutative space. Finally, we considered a Lie 2-algebra that is isomorphic to that obtained from the quantization of $\FR^3$, to demonstrate what is to be expected for a more general categorified correspondence principle.

Altogether, we conclude that Lie 2-algebra models are generalizations of the IKKT model that contain various other zero-dimensional models that were proposed in the context of M2-brane models. Moreover, many of the nice features of the IKKT model carry over to these Lie 2-algebra models.

One of our original motivations for studying Lie 2-algebra models was to explore the possibility of supersymmetric such models. This is particularly interesting, as there is more and more evidence that M2- and M5-brane models should be based on semistrict Lie 2-algebras, see e.g.\ \cite{Palmer:2012ya,Palmer:2013pka}. The reason for focusing on the zero-dimensional case instead of the three- and six-dimensional cases is that here the gauge structure severely simplifies. For a brief overview over the complications encountered in the higher dimensional case, see appendix \ref{app:B}. The construction of supersymmetric Lie 2-algebra models corresponding to a dimensional reduction of six-dimensional superconformal models is clearly an issue that we plan to attack in future work. Moreover, recall that the IKKT model was connected to type IIB superstring theory via a Schwinger-Dyson equation for the Wilson loops \cite{Fukuma:1997en}. It would be very interesting to study the corresponding equations for Wilson surfaces in our Lie 2-algebra models. 

Further open questions arising from our work concern a potential use of Lie 2-algebras in the regularization of Nambu-Poisson sigma-models as well as the development of our na\"ive notion of quantization of 2-plectic manifolds to a full quantization. The latter problem would imply to extend our Lie 2-algebras to Poisson 2-algebras or, equivalently, Gerstenhaber algebras, in which also a categorified associative product between observables is realized.

\section*{Acknowledgements}

We would like to thank Cedric Troessaert for discussions and the anonymous referee for an extraordinarily helpful report. PR was supported by Fondecyt grant \#3120077. CS was supported by an EPSRC Career Acceleration Fellowship. The Centro de Estudios Cientf\'icos (CECS) is funded by the Chilean Government through the Centers of Excellence Base Financing Program of Conicyt. 

\appendices

\subsection{Useful definitions}\label{app:A}

\paragraph{Strong homotopy Lie algebras.} An {\em $L_\infty$-algebra} or {\em strong homotopy Lie algebra} is a graded vector space $L=\oplus_i L_i$ endowed with $n$-ary multilinear totally antisymmetric products $\mu_n$, $n\in\NN^*$, of degree $2-n$, that satisfy homotopy Jacobi identities, cf.\ \cite{Lada:1992wc,Lada:1994mn,0821843621}. These identities read as
\begin{equation}\label{eq:homotopyJacobi}
 \sum_{i+j=n}\sum_\sigma\chi(\sigma;x_1,\ldots,x_n)(-1)^{i\cdot j}\mu_{j+1}(\mu_i(x_{\sigma(1)},\cdots,x_{\sigma(i)}),x_{\sigma(i+1)},\cdots,x_{\sigma(i+j)})=0
\end{equation}
for all $n\in \NN^*$, where the sum over $\sigma$ is taken over all $(i,j)$ unshuffles. Recall that a permutation $\sigma$ of $i+j$ elements is called an {\em $(i,j)$-unshuffle}, if the first $i$ and the last $j$ images of $\sigma$ are ordered: $\sigma(1)<\cdots<\sigma(i)$ and $\sigma(i+1)<\cdots<\sigma(i+j)$. Moreover, the {\em graded Koszul sign} $\chi(\sigma;x_1,\cdots,x_n)$, $x_i\in L$ is defined via the equation
\begin{equation}
 x_1\wedge \cdots \wedge x_n=\chi(\sigma;x_1,\cdots,x_n)\,x_{\sigma(1)}\wedge \cdots \wedge x_{\sigma(n)}
\end{equation}
in the free graded algebra $\wedge (x_1,\cdots,x_n)$, where $\wedge$ is considered graded antisymmetric.

Note that for elements of $L$ which do not have a definite grading, the above relations have to be resolved to elements of $L$ with homogeneous grading, using linearity of the maps. Note also that we shall denote the grading of an object $x$ by $\xt\in \RZ$. For example, we have $\xt=i$ for $x\in L_i$.

Strong homotopy Lie algebras that are concentrated in degrees $-n+1,\ldots,0$, i.e.\ $L_i=\varnothing$ for $i\notin [-n+1,\ldots,0]$, are categorically equivalent to semistrict Lie $n$-algebras.

\paragraph{Nambu-Poisson structures.} A {\em Nambu-Poisson structure}~\cite{Nambu:1973qe,Takhtajan:1993vr}
on a smooth manifold $\CM$ is an $n$-ary, totally antisymmetric multilinear map $\{\dotsp,\dots ,\dotsp\}:\CC^\infty(\CM)^{\wedge n}\rightarrow\CC^\infty(\CM)$, which satisfies the {\em generalized Leibniz rule}
\begin{equation}
 \{f_1 \,f_2,f_3,\dots ,f_{n+1}\}=f_1\,\{f_2,\dots
 ,f_{n+1}\}+\{f_1,\dots ,f_{n+1}\} \,f_2
\end{equation}
as well as the {\em fundamental identity}
\begin{equation}
 \{f_1,\dots ,f_{n-1},\{g_1,\dots ,g_n\}\}=\{\{f_1,\dots ,f_{n-1},g_1\},\dots ,g_n\}+\dots +\{g_1,\dots ,\{f_1,\dots ,f_{n-1},g_n\}\}
\end{equation}
for all $f_i,g_i\in\CC^\infty(\CM)$. A manifold $M$ endowed with such a {\em Nambu $n$-bracket} giving rise to a {\em Nambu-Poisson algebra} is called a {\em Nambu-Poisson manifold}. Under certain conditions, 2-plectic structures give rise to ternary Nambu-Poisson structures \cite{springerlink:10.1007/BF00400143}.

\paragraph{$n$-Lie algebras.} An $n$-Lie algebra\footnote{which is not to be confused with a Lie $n$-algebra arising in the categorification of Lie algebras} \cite{Filippov:1985aa} is a vector space $\CA$ endowed with an $n$-ary, totally antisymmetric and multilinear map $[\dotsp,\cdots,\dotsp]:\CA^{\wedge n}\rightarrow \CA$ that satisfies the {\em fundamental identity}:
\begin{equation}\label{eq:FI_algebra}
 [a_1,\dots ,a_{n-1},[b_1,\dots ,b_n]]=[[a_1,\dots ,a_{n-1},b_1],\dots ,b_n]+\dots +[b_1,\dots ,[a_1,\dots ,a_{n-1},b_n]]
\end{equation}
for all $a_i,b_i\in\CA$. Note that Nambu-Poisson algebras are particular $n$-Lie algebras, and it has been proposed that Nambu-Poisson structures should be quantized in terms of $n$-Lie algebras, cf.\ \cite{DeBellis:2010pf} and references therein.

Note that $n$-Lie algebras come with a Lie algebra of inner derivations, which are given by linear combinations of the maps
\begin{equation}
 D(a_1,\ldots,a_{n-1})\acton x:=[a_1,\ldots,a_{n-1},x]~,
\end{equation}
where $a_i,x\in \CA$. The commutator of inner derivations closes on inner derivations because of the fundamental identity \eqref{eq:FI_algebra}.

We can endow an $n$-Lie algebra with a metric, which has to be invariant under the action of inner derivations. In the case of a 3-Lie algebra, this metric induces a metric on the vector space of inner derivations, which is in general indefinite and different from the Killing form.

The first of the recently studied M2-brane models, the Bagger-Lambert-Gustavsson (BLG) model \cite{Bagger:2007jr,Gustavsson:2007vu}, has a gauge structure that is based on a 3-Lie algebra. This 3-Lie algebra comes with a positive definite metric on the vector space forming the 3-Lie algebra, which induces a metric of split signature on the inner derivations.

\paragraph{Generalized 3-algebras.} Because there is essentially only one finite-dimensional 3-Lie algebras with positive definite invariant metric, various generalizations have been proposed. First, there are the hermitian 3-algebras that are based on a complex vector space and that underlie the ABJM M2-brane model \cite{Aharony:2008ug,Bagger:2008se}. Second, there are the real 3-algebras, which are relaxed versions of 3-Lie algebras in that their 3-bracket is antisymmetric only in the first two slots \cite{Cherkis:2008qr}. Both types of 3-algebras can be encoded in terms of Lie algebras and certain representations \cite{deMedeiros:2008zh}, and therefore they form differential crossed modules \cite{Palmer:2012ya}. Also, as shown in the text, skeletal Lie 2-algebras with inner product come naturally with a generalized 3-algebra structure.

\subsection{Gauge symmetry in semistrict higher gauge theory}\label{app:B}

While semistrict higher gauge theory has only been developed partially, an attempt to capture its local gauge structure has been made in \cite{Zucchini:2011aa}. Below, we will give a rough, quick review of this construction. This serves two purposes. First, we can easily show that it reduces to the Lie 2-algebra homomorphisms describing the symmetries of our Lie 2-algebra models. Second, it demonstrates that it is considerably simpler to study Lie 2-algebra models than to study actual semistrict higher gauge theories.

Let us group the fields we are interested in working with into
doublets $(\phi,\Phi)\in\Omega^p(M,W)\times\Omega^{p+1}(M, V)$, where
$M$ indicates the manifold they live on and small and capital letters
will always indicate $V$- or $W$-valued fields respectively. We will
refer to the \textit{degree} of the doublet as the order of the
$W$-valued form, in this case $p$. Let us
indicate a \textit{connection doublet} by
$(a,A)\in\Omega^1(M,W)\times\Omega^2(M,V)$. We can define the
curvature of these fields as the doublet
$(f,F)\in\Omega^2(M,W)\times\Omega^3(M,V)$:
\begin{align}
  \label{eq:29}
  f=&\dd a+\tfrac{1}{2}\mu_2(a,a)-\mu_1(A)~,\\
F=&\dd A+\mu_2(a,A)-\tfrac{1}{6}\mu_3(a,a,a)~.
\end{align}
The $(f,F)$ doublet can be easily seen to satisfy the \textit{Bianchi
  identities}
\begin{align}
  \label{eq:30}
 \dd f+\mu_2(a,f)+\mu_1(F)=&0~,\\
\dd F+\mu_2(a,F)-\mu_2(f,A)+\tfrac{1}{2}\mu_3(a,a,f)=&0~.
\end{align}
In analogy to ordinary gauge theory, one would like the Bianchi
identities to be given by the requirement that $Df=0=DF$, where $D$ is
the \textit{covariant derivative} with respect to the same connection $(a,A)$. This requirement allows one to
define the action of $D$ on a generic field doublet $(\phi,\Phi)$ of
order $p$ as
\begin{align}
  D\phi&=\dd\phi+\mu_2(a,\phi)+(-1)^p\mu_1(\Phi)~,\label{eq:31}\\
D\Phi&=\dd\Phi+\mu_2(a,\Phi)-(-1)^p\mu_2(\phi,A)+\frac{(-1)^p}{2}\mu_3(a,a,\phi)\label{eq:32}~,
\end{align}
forming the $(p+1)$-degree doublet $(D\phi,D\Phi)$.
The next step is to define gauge transformations in the
semistrict Lie 2-algebra setting. These have to live in the set of
automorphisms of the Lie 2-algebra and are expected to satisfy a generalization of the Maurer-Cartan equation $\dd(g^{-1}\dd g)+(g^{-1}\dd g)\wedge(g^{-1}\dd g)=0$. It is argued in \cite{Zucchini:2011aa}
that the easiest way to generalize traditional gauge theory also
makes use of a flat connection doublet $(\sigma,\Sigma)$, which
roughly speaking keeps track of how gauge group elements vary with
respect to the base manifold coordinates. Overall,  the semistrict Lie 2-algebra 1-gauge transformations are
defined in \cite{Zucchini:2011aa} as the following set of ingredients:
\begin{conditions}
\item[(i)] a map $g\in\text{Map}(M,\sAut(L))$, i.e. a set
  $(g_0,g_{-1},g_2)$ satisfying the requirements elucidated in section
  \ref{ssec:Lie_2_algebra_homomorphism}~;
\item[(ii)] a flat connection doublet $(\sigma,\Sigma)$:
\begin{subequations}\label{eq:semistict_gauge_consistency}
  \begin{align}
    \label{eq:33}
    \dd\sigma+\tfrac{1}{2}\mu_2(\sigma,\sigma)-\mu_1(\Sigma)=&0~,\\
\dd\Sigma +\mu_2(\sigma,\Sigma)-\tfrac{1}{6}\mu_3(\sigma,\sigma,\sigma)=&0~;
  \end{align}
\item[(iii)] an element $\tau\in\Omega^1(M,\sHom(W,V))$ satisfying
  \begin{equation}
    \label{eq:35}
    \dd\tau(w)+\mu_2(\sigma,\tau(w))-\mu_2(w,\Sigma)+\tfrac{1}{2}\mu_3(\sigma,\sigma,w)+\tau\left(\mu_2(\sigma,w)+\mu_1(\tau(w))\right)=0~.
  \end{equation}
\end{subequations}
\end{conditions}
\vspace{-0.5cm}
Note that equation~\eqref{eq:33} is just the vanishing fake curvature
condition as we know it from strict higher gauge theory, while
equation~\eqref{eq:35} is referred to as the \textit{2-Maurer-Cartan equation} in \cite{Zucchini:2011aa}. Indeed, after
introducing a $\tau$-dependent term in the definition of the action of
the flat connection, for instance
\begin{align}
  \label{eq:36}
  g_0^{-1}\dd g_0(w)-\mu_2(\sigma,w)-\mu_1(\tau(w))=0~
\end{align}
for the $W$ part of the automorphism $g$, one can satisfy the $W$ sector
of the Maurer-Cartan equation (and analogously for the $V$-part). Because of the non-vanishing Jacobiator in the
semistrict set-up, without $\tau$ this is normally not possible
unless one imposes a further condition by hand. \\
Apart from equation~\eqref{eq:36} and its $V$-sector analogue, there is a further
condition of compatibility for $\tau$, so that the set of homomorphism
rules of section~\ref{ssec:Lie_2_algebra_homomorphism} are still
satisfied for $g$ - the interested reader can find all the details for
this construction in \cite{Zucchini:2011aa}.

Now, to see how the above defined gauge transformations act on fields,
one requires that, for a given connection doublet $(a,A)$, the
covariant derivatives $D$ from~\eqref{eq:31} and~\eqref{eq:32} ``pull through'' all the
elements that make up the transformation. That is, if we indicate the
full 1-gauge transformation by $(g,\sigma,\Sigma,\tau)$, one requires the
derivatives $D$ to act on gauge transformed field doublets
$(g\acton\phi,g\acton\Phi)$ as defined in \eqref{eq:31} and \eqref{eq:32}, but treating all the components $g,\,\sigma,\,\Sigma$
and $\tau$ as covariantly constant. In this way one obtains for
connection doublets $(a,A)$ the following action of 1-gauge transformations:
\begin{align}
  \label{eq:37}
  g\acton a=&g_0(a-\sigma)~,\\
g\acton A=&g_{-1}(A-\Sigma+\tau(a-\sigma))-\tfrac{1}{2}g_2(a-\sigma,a-\sigma)~,
\end{align}
so that its covariant derivative, or curvature doublet $(f,F)\equiv
(Da,DA)$ transforms as
\begin{align}
  \label{eq:20}
  g\acton f =&g_0(f)~,\\
g\acton F=&g_{-1}(F-\tau(f))+g_2(a-\sigma,f)~.
\end{align}
Similarly one can define \textit{canonical} field doublets
$(\phi,\Phi)$, of degree $p$, as those that transform as
\begin{align}
  \label{eq:38}
  g\acton\phi=&g_0(\phi)~,\\
g\acton\Phi=&g_{-1}(\Phi-(-1)^p\tau(\phi))+(-1)^pg_2(a-\sigma,\phi)~.
\end{align}
Its covariant derivatives then transform as
\begin{align}
  \label{eq:61}
  g\acton D\phi=&g_0(D\phi)~,\\
g\acton D\Phi=&g_{-1}(D\Phi+(-1)^p\tau(D\phi))-(-1)^pg_2(a-\sigma,D\phi)+(-1)^pg_2(Da,\phi)~.
\end{align}
The obvious thing to notice here is that while in the $W$ sector
everything transforms in a ``nice'' way, that is $\phi\rightarrow
g_0(\phi)$ and $D\phi\rightarrow g_0(D\phi)$, the $V$ sector looks a
lot more involved. When constructing actions, these will be based on
some inner product that will be invariant under 
automorphisms of the Lie 2-algebra and therefore under the transformation corresponding to
$(g_0,g_{-1})$. In this sense it would be very easy to identify and
construct gauge invariant actions if covariant derivatives and
curvatures transformed as $D\Phi\rightarrow g_{-1}(D\Phi)$ also in the
$V$ sector of actions. Interestingly this can be achieved: the
Lie 2-algebra can be \textit{gauge rectified}\footnote{It has not been
  shown whether a pair $(\lambda,\rho)$ of gauge rectifiers can always
  be found, for any Lie 2-algebra.} by a pair of fields
$(\lambda, \rho)$, where $\lambda\in\Omega^0(M,\sHom(W\wedge
W,V))$ and $\rho\in\Omega^1(M,\sHom(W,V))$, which have special
gauge transformation properties. The products $\mu_2$ and $\mu_3$ can
then be corrected by $\lambda$ so that the rectified products $\mu_i^{(\lambda)}$
will transform as
$g\acton\mu_i^{(\lambda)}(\ldots)=g_\alpha(\mu_i^{(\lambda)}(\ldots))$, where
$\alpha=0,-1$ according to where $\mu_i$ maps to. Similarly field doublets
can be rectified, as well as the definition of the covariant
derivative, resulting in all fields and covariant derivatives thereof
transforming simply by $g_0$ or $g_{-1}$ actions on the objects
themselves (e.g. $D^{(\lambda,\rho)}\Phi^{(\lambda,\rho)}\rightarrow g_{-1}(\mu_i^{(\lambda)}(\ldots))$). This means that for
actions constructed via a $g_0,\,g_{-1}$ invariant inner product, any
terms involving rectified canonical field doublets, covariant
derivatives thereof and Lie 2-algebra products $\mu^{(\lambda)}_i$
will be automatically gauge invariant. Moreover, the rectified
products $\mu_i^{(\lambda)}(\ldots)$ still form a Lie 2-algebra. To
see the details of this procedure we again refer to
\cite{Zucchini:2011aa}.

Returning to the gauge transformation setup, upon reduction to zero
dimensions all the total derivatives disappear and therefore the
auxiliary $\sigma,\,\Sigma$ and $\tau$ can all be set to zero. Also, we
cannot talk about field doublets anymore, since all objects we will be
considering are of order 0, whether they are valued in $W$ or in
$V$. This simplifies matters considerably, as we can now say that the
gauge transformations of $w\in W$ and $v\in V$ are given by
\begin{equation}
  \label{eq:69}
  g\acton w=g_0(w) \qquad \text{and}\qquad g\acton v=g_{-1}(v)~,
\end{equation}
while covariant derivation reduces to
\begin{equation}
  \label{eq:70}
  D w=\mu_2(a,w) \qquad\text{and}\qquad D v=\mu_2(a,v)~,
\end{equation}
for a 0-degree field $a\in W$. We set $(g_0,g_{-1},g_2)$ as in
\ref{ssec:actions_and_symmetries} to:
\begin{equation}
  \label{eq:71}
  g_0(w):=w+\mu_2(\epsilon,w)~,\quad g_{-1}(v):=v+\mu_2(\epsilon,v)~, \quad g_2(w_1,w_2):=\mu_3(\epsilon,w_1,w_2)~,
\end{equation}
to first order in the gauge parameter $\epsilon\in W$. It then follows from the
homomorphism rules that
covariant derivatives transform in the desired way:
\begin{align}
  \label{eq:72}
  g\acton D w=&g_0(Dw)=\mu_2(g\acton a, g\acton w)=\mu_2(g_0(a),g_0(w))~,\\
  g\acton D v=&g_1(Dv)=\mu_2(g\acton a, g\acton v)=\mu_2(g_0(a),g_{-1}(v))~,
\end{align}
as expected. Indeed, the homomorphism rules themselves guarantee that
all the
2-algebra products on $w,\,v$ also transform simply by an overall
$g_\alpha$, that is $\mu_i(\ldots)\rightarrow g_\alpha\mu_i(\ldots)$, for
$\alpha=0,-1$ according to the grading of $\mu_i$. In other words, for
zero-dimensional reduced actions, if based on a $\mu_2(w,\dotsp)$ invariant
inner product, any terms involving the 2-algebra structures are
automatically gauge invariant, without the need to introduce any
rectifiers.

\bibliographystyle{latexeu}

\bibliography{bigone}

\end{document}